\documentclass[longauth]{aa}  

\usepackage{graphicx}
\usepackage{txfonts}

\usepackage{lscape}
\usepackage{hyperref}
\usepackage{xcolor}
\usepackage{gensymb}
\usepackage{natbib}
\usepackage{multirow}
\usepackage[capitalize]{cleveref}
\usepackage{ulem}

\hypersetup{
	pdfnewwindow=true,  
	colorlinks=true,    
	linkcolor=blue,  
	citecolor=blue,    
	filecolor=cyan,     
	urlcolor=black,      
} 

\begin{document}

\title{Behind the dust veil: A panchromatic view of an optically dark galaxy at z=4.82}

\titlerunning{Behind the dust veil}
\authorrunning{Sillassen et al.}

\author{
Nikolaj B. Sillassen\inst{1,2},
Shuowen Jin\inst{1,2,\thanks{Marie Curie Fellow}},
Georgios E. Magdis\inst{1,2,3},
Jacqueline Hodge\inst{4},
Raphael Gobat\inst{5},
Emanuele Daddi\inst{6},
Kirsten Knudsen\inst{7},
Alexis Finoguenov\inst{8},
Eva Schinnerer\inst{9},
Wei-Hao Wang\inst{10},
Zhen-Kai Gao\inst{10},
John R. Weaver\inst{11},
Hiddo Algera\inst{10,12,13},
Irham T. Andika\inst{14,15},
Malte Brinch\inst{16},
Chian-Chou Chen\inst{17},
Rachel Cochrane\inst{18,19},
Andrea Enia\inst{20,21},
Andreas Faisst\inst{22},
Steven Gillman\inst{1,2},
Carlos Gomez-Guijarro\inst{6},
Ghassem Gozaliasl\inst{8,23},
Chris Hayward\inst{24},
Vasily Kokorev\inst{25},
Maya Merchant\inst{1,3},
Francesca Rizzo\inst{26},
Margherita Talia\inst{27,21},
Francesco Valentino\inst{1,2},
David Blánquez-Sesé\inst{1,2},
Anton M. Koekemoer\inst{28},
Benjamin Magnelli\inst{6},
Michael Rich\inst{29},
Marko Shuntov\inst{1,3}
          }
   \institute{Cosmic Dawn Center (DAWN), Copenhagen, Denmark\\
      \email{nbsi@space.dtu.dk}
    \and
            DTU-Space, Technical University of Denmark, Elektrovej 327, DK-2800 Kgs. Lyngby, Denmark\\
        \email{shuji@dtu.dk}
    \and
            Niels Bohr Institute, University of Copenhagen, Jagtvej 128, DK-2200 Copenhagen, Denmark
    \and
            Leiden Observatory, Leiden University, NL-2300 RA Leiden, the Netherlands
    \and
            Instituto de Física, Pontificia Universidad Católica de Valparaíso, Casilla 4059, Valparaíso, Chile
    \and
            Université Paris-Saclay, Université Paris Cité, CEA, CNRS, AIM, Paris, France
    \and
            Department of Space, Earth, \& Environment, Chalmers University of Technology, Chalmersplatsen 4 412 96 Gothenburg, Sweden
    \and
            Department of Physics, University of Helsinki, PO Box 64, 00014 Helsinki, Finland 
    \and
            Max-Planck-Institut für Astronomie, Königstuhl 17, D-69117 Heidelberg, Germany
    \and
            Academia Sinica Institute of Astronomy and Astrophysics (ASIAA), No. 1, Section 4, Roosevelt Rd., Taipei 106216, Taiwan
    \and
            Department of Astronomy, University of Massachusetts, Amherst, MA 01003, USA
    \and
            Hiroshima Astrophysical Science Center, Hiroshima University, 1-3-1 Kagamiyama, Higashi-Hiroshima, Hiroshima 739-8526, Japan
    \and
            National Astronomical Observatory of Japan, 2-21-1, Osawa, Mitaka, Tokyo, Japan
    \and
            Technical University of Munich, TUM School of Natural Sciences, Department of Physics, James-Franck-Str. 1, D-85748 Garching, Germany
    \and
            Max-Planck-Institut für Astrophysik, Karl-Schwarzschild-Str. 1, D85748 Garching, Germany
    \and
            Instituto de Física y Astronomía, Universidad de Valparaíso, Avda. Gran Bretana 1111, Valparaíso, Chile
    \and
            Academia Sinica Institute of Astronomy and Astrophysics (ASIAA), No. 1, Section 4, Roosevelt Road, Taipei 106216, Taiwan
    \and
            Institute for Astronomy, University of Edinburgh, Royal Observatory, Blackford Hill, Edinburgh, EH9 3HJ, UK
    \and
            Department of Astronomy, Columbia University, New York, NY 10027, USA
    \and
            University of Bologna– Department of Physics and Astronomy “Augusto Righi” (DIFA), Via Gobetti 93/2, I-40129 Bologna, Italy
    \and
            INAF–Osservatorio di Astrofisica e Scienza dello Spazio, Via Gobetti 93/3, I-40129 Bologna, Italy
    \and
            Caltech/IPAC, 1200 E. California Boulevard, Pasadena, CA 91125, USA
    \and
            Department of Computer Science, Aalto University, PO Box 15400, Espoo 00 076, Finland
    \and
            Center for Computational Astrophysics, Flatiron Institute, 162 Fifth Avenue, New York, NY 10010, USA
    \and
            Department of Astronomy, The University of Texas at Austin, Austin, TX 78712, USA
    \and
            Kapteyn Astronomical Institute, University of Groningen, Landleven 12, 9747 AD, Groningen, The Netherlands
    \and
            University of Bologna, Department of Physics and Astronomy (DIFA), Via Gobetti 93/2, I-40129 Bologna, Italy
    \and
            Space Telescope Science Institute, 3700 San Martin Drive, Baltimore, MD 21218, USA
    \and
            Department of Physics \& Astronomy, University of California Los Angeles, 430 Portola Plaza, Los Angeles, CA 90095, USA
             }
   \date{Received XX / Accepted XX}

 \abstract
{
Optically dark dusty star-forming galaxies (DSFGs) play an essential role in massive galaxy formation at early cosmic time, however their nature remains elusive.
Here we present a detailed case study of all the baryonic components of a $z=4.821$ DSFG, XS55. Selected from the ultra-deep COSMOS-XS 3GHz map with a red SCUBA-2 450$\mu$m/850$\mu$m colour, XS55 was followed up with ALMA 3mm line scans and spectroscopically confirmed to be at $z=4.821$ via detections of the CO(5-4) and [C{\tiny I}](1-0) lines. 
JWST/NIRCam imaging reveals that XS55 is a F150W-dropout with red F277W/F444W colour, and a complex morphology: a compact central component embedded in an extended structure with a likely companion. 
XS55 is tentatively detected in X-rays with both Chandra and XMM-Newton, suggesting an active galactic nucleus (AGN) nature. By fitting a panchromatic SED spanning NIR to radio wavelengths, we revealed that XS55 is a massive main-sequence galaxy with a stellar mass of $M_\ast=(5\pm1)\times10^{10}\,{\rm M_\odot}$ and a star formation rate of ${\rm SFR}=540\pm177~{\rm M_\odot\,yr^{-1}}$. The dust of XS55 is optically thick in the far infrared (FIR) with a surprisingly cold dust temperature of $T_{\rm dust}=33\pm2\,{\rm K}$, making XS55 one of the coldest DSFGs at $z>4$ known to date. This work unveils the nature of a radio-selected F150W-dropout, suggesting the existence of a population of DSFGs hosting active black holes embedded in optically thick dust.
}

\maketitle
 

\section{Introduction}

Optically faint/dark galaxies are a population of massive dusty star-forming galaxies (DSFGs) that are faint or undetected in deep optical images while bright at longer wavelengths. 
This population includes but is not limited to the following samples: sub-millimetre galaxies (SMGs) \citep[e.g.,][]{Walter2012}, H-band dropouts \citep[e.g.,][]{Wang2019Natur,Pampliega2019,Smail2023}, Ks-faint \citep[e.g.,][]{Smail2020}, Hubble Space Telescope (HST)-dark/faint galaxies \citep[e.g.,][]{Franco2018,Perez-Gonzalez2023,Xiao2023_OFGs,Gomez-Guijarro2023}, and radio-selected near infrared (NIR) dark galaxies (\citealt{Algera2020,Talia2021,Enia2022,vanderVlugt2023_COSMOS_XS,Gentile2024a,Gentile2024}).
Recent studies revealed that this population contributes significantly ($10-40\%$) to the cosmic star formation rate density in the early universe ($z\sim3-6$, \citealt{Wang2019Natur,Fudamoto2021,Talia2021,Enia2022,Shu2022,Xiao2023_OFGs}), and dominates the massive end of the stellar mass function (SMF) at $z\sim3-8$ \citep{Wang2019Natur,Gottumukkala2024_OFG_SMF}. This indicates that this population plays a significant role in cosmic star formation history, even up to 50\% estimated from Lyman break galaxy selected samples at $z\sim3$ \citep{Enia2022}, but has been largely missed by previous optical/NIR surveys. 
Thanks to the unprecedented sensitivity and long wavelength coverage, the {\it James Webb Space Telescope} (JWST) can efficiently detect these objects and allows for detailed studies and large sample census of optically faint galaxies. 
For example, \cite{Barrufet2023} studied the stellar emission of a sample of 30 HST-dark sources in the CEERS field with red colours through 1.6\,--\,4.4\,${\rm \mu}$m in JWST NIRCam filters, revealing them all to be heavily dust obscured massive main sequence galaxies and significantly contributing to the star formation rate density at high redshift. 
In a NIRSpec study of 23 HST-dark galaxies, \citet{Barrufet2024_OFGs} found the majority to be massive ($\log(M_\ast/\mathrm{M_\odot})>9.8$) and highly attenuated ($A_V>2$) star-forming galaxies with a broad range of recent star-formation activities.
\cite{Kokorev2023_dustyguy} studied the multi-wavelength properties of an HST-dark galaxy at $z_{\rm spec}=2.58$ with JWST/NIRCam data, revealing it to be a massive disk galaxy with log$(M_{\star}/\mathrm{M_\odot})>11$ and optically thick dust in the far-infrared. They also found that such objects would not be detected in JWST filters bluer than F356W if placed at $z>4$, and becoming JWST-dark at $z>6$ at current depths of major JWST surveys.
Given their extreme faintness in optical and NIR wavelengths, and brightness in far infrared (FIR), (sub)mm facilities like ALMA and NOEMA, are more efficient in confirming their redshifts via detecting CO and/or [C{\tiny I}] lines \citep[e.g.,][]{Weiss2009,Walter2012,Riechers2013Nature,Riechers2017,Jin2019alma,Jin2022,Casey2019,Birkin2021,Chen2022,Gentile2024}, which hence revealed vigorous star-bursting activities with obscured star formation rate ${\rm SFR}\sim300-3000\,{\rm M_\odot/yr}$ and large gas reservoirs $\log M_{\rm gas}/{\rm M_\odot}\sim 10.5-11.5$ in these massive systems \citep[e.g.,][]{Riechers2013Nature,Jin2022}. However, the spectroscopic sample is still small and strongly biased towards sources with the brightest submm fluxes.
Consequently, studies of this population strongly rely on photometric redshifts, which entail the risk of catastrophic failures. 
For example, \citet{Ling2024_cosbo7} reported an optically dark ${\rm photo-}z>7$ DSFG candidate using 10 bands of JWST photometry, however, it was eventually confirmed at ${\rm spec-}z=2.625$ by \cite{Jin2024_cosbo7} via multiple CO and [CI] line detections. 
Evidently, robust spectroscopic confirmation is essential to uncover the nature of these extreme dusty objects.

Despite the limited sample of spectroscopically confirmed DSFGs, recent studies have revealed optically faint/dark DSFGs have optically thick dust in FIR, massive gas reservoirs and elevated star formation efficiency (SFE) compared to main sequence galaxies \citep{Jin2019alma,Jin2022,Kokorev2023_dustyguy}. Nevertheless, it remains unclear whether active galactic nuclei (AGN) are present in these systems. As both X-ray and optical AGN features can be severely attenuated by dust, the AGN fraction of DSFGs could be largely underestimated \citep{Franco2018}. Therefore, panchromatic studies including deep radio observations \citep[e.g,][]{Delvecchio2017,Algera2020,van_der_Vlugt2021_COSMOS_XS,vanderVlugt2023_COSMOS_XS,Gentile2024} are key to identify potential AGN activity in these systems. As previous studies focus either on the stellar or interstellar medium (ISM) components, comprehensive studies of all baryonic components (stellar, dust, and gas) and AGN activity are essential to unveil their nature.

In this letter, we report the spectroscopic confirmation of the radio-selected DSFG XS55 and provide a panchromatic view of its stellar, dust, and gas components and associated AGN activity.
We adopt a flat cosmology with $H_0=70\,{\rm km\,s^{-1}\,Mpc^{-1}}$, $\Omega_{\rm M}=0.27$, and $\Omega_\Lambda=0.73$, and use a \citet{Chabrier2003} initial mass function (IMF). All magnitudes are in the AB system \citep{Oke1974_AB_sys}.

\section{Selection and Data}
\subsection{Selection}
XS55 was originally selected in the COSMOS-XS \citep{Algera2020,van_der_Vlugt2021_COSMOS_XS} catalog with an ID=55, hence we dub it XS55. It is detected in the ultra-deep COSMOS-XS S-band image with $S_{\rm 3GHz}$ = $6.35\pm0.96\,{\rm \mu Jy}$, but undetected in the less deep COSMOS 3\,GHz map \citep[${\rm rms=2.5\,\mu Jy}$,][]{Smolcic2017}, and detected in the MeerKAT image with $S_{{\rm 1.3GHz}}=10.9\pm2.1\,{\rm \mu Jy}$  \citep{Jarvis2016mightee,Heywood2022,Hale2024_meerkat}.
XS55 has no optical counterpart (i.e., optically dark) and is not included in the COSMOS2020 catalogue \citep{Weaver2022COSMOS2020}. It drops out in JWST F115W and F150W images, but is detected in IRAC 4.5$\mu$m (\cref{fig:cutouts}) and tentatively detected ($\sim3\sigma$) in the ALMA 2\,mm MORA map (\citealt{Casey2021_MORA}).
By performing the super-deblending technique \citep{Jin2018cosmos,Liu_DZ2017} with the radio prior, we measure the deblended {\it Herschel} and SCUBA-2 photometry of XS55. Interestingly, it is not detected in {\it Herschel} images ($3\sigma$ limiting depths: 250\,${\rm \mu}$m=$5.3\, {\rm mJy}$,350\,${\rm \mu}$m=$8.0\,{\rm mJy}$, 500\,${\rm \mu}$m=$8.7\,{\rm mJy}$, \citealt{Jin2018cosmos}), but is well detected in two SCUBA-2 bands; 450$\mu$m \citep{Gao2024_SCUBA450} and 850$\mu$m \citep{Simpson2019} with a red 450$\mu$m/850$\mu$m colour ($S_{\rm450\,\mu m}=5.6\pm1.4\,{\rm mJy}, S_{\rm850\,\mu m}=5.7\pm0.8\,{\rm mJy}$). Assuming typical dust templates from \citet{Magdis2012SED}, the red SCUBA-2 colour suggests a FIR photometric redshift of $z>6$. Consequently, XS55 was followed up by two ALMA 3\,mm line scan projects in Cycle 9 (ID: 2022.1.00884, PI: R. Gobat; ID: 2022.1.00863.S, PI: J. Hodge).

\subsection{ALMA}
The two ALMA programmes were observed for a total of 1.9 hours on source.
The frequency setups are identical in the two programs, adopting the same setups as in \cite{Jin2019alma} and covering 84--108 GHz with three tunings. We produce measurement sets with the Common Astronomy Software Applications (CASA, \citealt{McMullin2007CASA}) pipeline for each observation programme.
Following the methods presented in \citet{Jin2022}, \citet{Zhou2024}, and \citet{Sillassen2024}, the calibrated data are converted to $uv$ table format and analysed with the \texttt{GILDAS} software package in $uv$ space. 
To enhance the signals, the $uv$ tables from the two programs are combined using the task $uv\_merge$. 
The final products reach a continuum sensitivity of 9.6$\,{\rm \mu Jy/beam}$ with a spatial resolution of $1\farcs6$, and a line sensitivity of 12$\,{\rm mJy/beam}$ over 500 km/s at $\sim 99\,{\rm GHz}$. As shown in see \cref{fig:spectrum,sec:redshift}, We robustly detect continuum ($\sim14\sigma$) and two lines ($\sim10\sigma$ and $\sim4\sigma$). The dust continuum is well-fitted by a point-source model using \texttt{GILDAS} \texttt{uvfit}, while fitting with an elliptical Gaussian does not yield useful constraints. Therefore, the dust continuum is unresolved, and we place an upper limit on the continuum size using Eq. (2) from  \citet{Gomez_Guijarro2022} (Table~\ref{tab:physpars}).

\subsection{JWST}
XS55 was observed with JWST NIRCam in F115W, F150W, F277W and F444W bands, as a part of the COSMOS-Web survey \citep{Casey2023_COSMOS_WEB}.  We use the image product versions from the Dawn JWST Archive \citep[DJA\footnote{\url{https://dawn-cph.github.io/dja/index.html}},][]{Valentino2023_DJA}, and further, have verified these are consistent with the COSMOS-Web team's map (Shuntov et al. in prep.). 
XS55 is well detected in both F277W ($\sim10\sigma$) and F444W ($\sim36\sigma$), but not detected in both F115W and F150W ($<3\sigma$), consistent with the H-dropout selection from \citet{Wang2019Natur}.

\subsection{X-ray}
\begin{figure*}[ht]
        \centering
        \includegraphics[width=\textwidth]{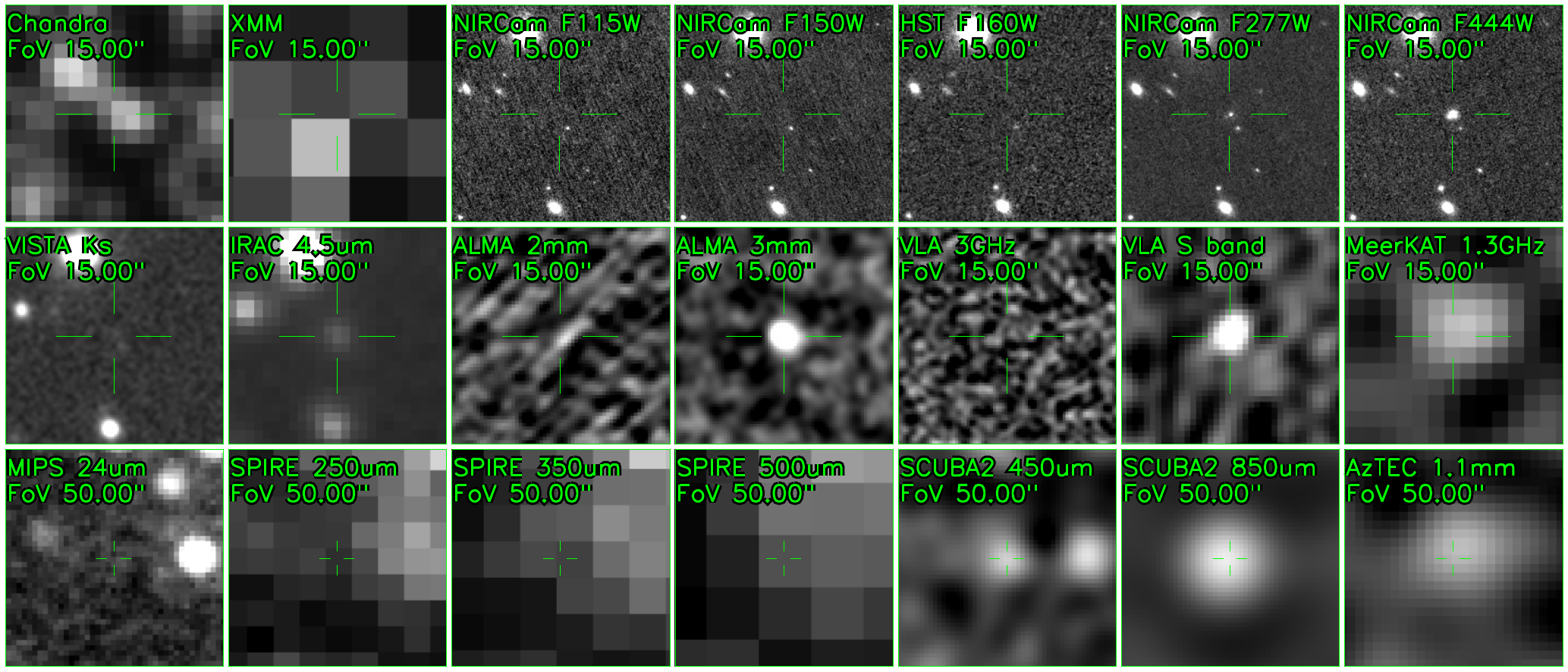}
        \caption{Multi-wavelength cutout images of XS55. The instrument, wavelength and field of view (FoV) are shown in green text in each panel.}
        \label{fig:cutouts}
\end{figure*}

The COSMOS field has been fully observed in soft (0.5-2.0\,{\rm keV}) and hard (2.0-10\,{\rm keV}) X-rays with both XMM-Newton \citep[50 ks per pointing, PI: G. Hasinger;][]{Hasinger2007_XMMCosmos} and Chandra ($\sim$180 ks exposure) as part of the Chandra COSMOS \citep[C-COSMOS, PI: M. Elvis;][]{Elvis2009_CCOSMOS} and Chandra COSMOS Legacy \citep[PI: F. Civano;][]{Civano2016} surveys. 
As shown in \cref{fig:color-img,fig:cutouts}, XS55 is tentatively detected in soft X-ray band of Chandra with $2.1\sigma$ significance, and detected in the stacked soft, medium and hard X-ray XMM-Newton images with $3.1\sigma$.

\section{Results}

\subsection{Redshift identification}
\label{sec:redshift}
\begin{figure*}[!htbp]
    \centering
    \includegraphics[width=\textwidth]{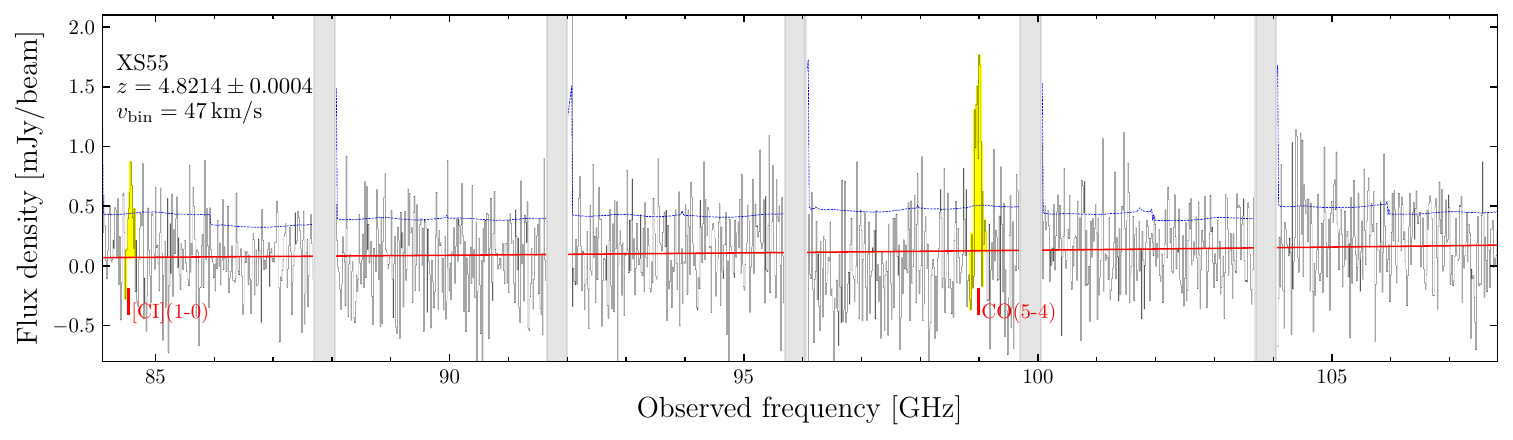}
    \raisebox{-0.577\height}{\includegraphics[width=0.284\textwidth]{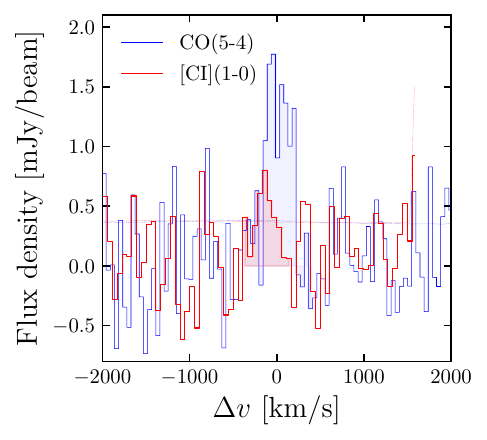}}
    \raisebox{-0.5\height}{\includegraphics[width=0.22\textwidth]{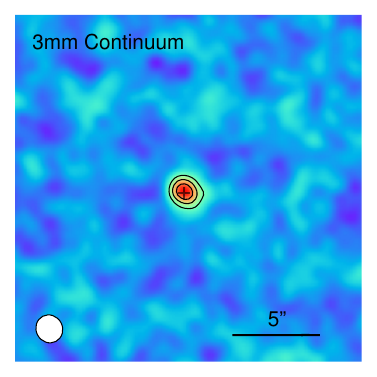}}
    \raisebox{-0.5\height}{\includegraphics[width=0.22\textwidth]{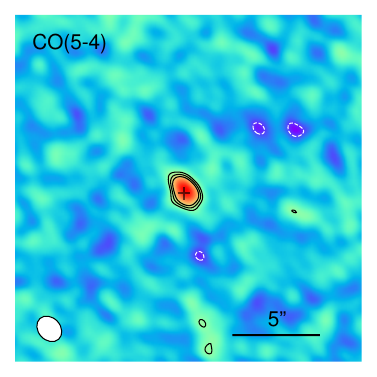}}
    \raisebox{-0.5\height}{\includegraphics[width=0.22\textwidth]{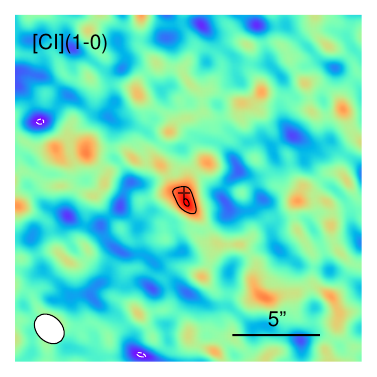}}
    \caption{{\bf Top:} ALMA 3mm spectrum of XS55. The red line shows the line-free continuum, and the blue dotted line indicates the flux error per channel at $1\sigma$ level. The spectroscopic redshift is shown in text, along with the velocity width of the channels. {\bf Bottom-left:} Velocity space spectrum of CO(5-4) (blue) and [C{\tiny I}](1-0) (red) at $z=4.8214$, the uncertainty per channel is shown as dashed lines. {\bf Bottom-right:} Continuum and continuum subtracted moment-0 line maps of XS55. Contour levels are 5, 8, and 11$\sigma$ for continuum, and 3, 4, 5$\sigma$ for the line maps. The beam size is shown as a white ellipse, and the peak JWST/F444W position is marked by a grey cross.}
    \label{fig:spectrum}
\end{figure*}

After extracting the ALMA spectra in all spectral windows (SPWs), we combine them in a single 1D spectrum (\cref{fig:spectrum}), and run a line-searching algorithm as in \citet{Jin2019alma} to search for emission line features with the highest significance. The continuum is fit with a power-law with fixed slope of 3.7 in frequency \citep[assuming $\beta\approx1.7$,][]{Magdis2012SED,Sillassen2024}, masking out the channels of significant emission lines. The detected emission lines are fitted with a Gaussian profile in the continuum subtracted spectrum. XS55 is detected in 3mm continuum at $\sim14\sigma$ with a flux of $134.2\pm9.3\,{\rm \mu Jy}$ at $\sim96\,{\rm GHz}$. One line is detected at $98.99\,{\rm GHz}$ at $9.8\sigma$ (see \cref{fig:spectrum}), and we search for other lines in the spectrum consistent with this detection. We find a $3.9\sigma$ detection at $84.58\,{\rm GHz}$. The two lines are consistent with CO(5-4) and [C{\tiny I}](1-0) at $z=4.8214\pm0.0004$. We note that there is a slight velocity offset between the CO and CI line peaks of $\sim141\,{\rm km/s}$ (\cref{fig:spectrum}). By defining the [CI] line range using the velocity range of the CO line, the [CI] S/N is $3.3\sigma$, yielding a low $P_{\rm chance}=0.4\%$, where $P_{\rm chance}$ is the chance probability of finding a spurious second line \citep{Jin2019alma}. Further, we compare the redshift solution with the NIR-SED fitted redshift probability distribution PDF(z) of \texttt{LePhare} in the COSMOS-Web catalog ($z_{\rm phot}=4.73^{+0.52}_{-0.64}$, Shuntov et al. in prep.), and find they are in excellent agreement (\cref{fig:full_sed}). For a sanity check, we tested the redshift solution of $z=3.66$ in the case of the bright line being CO(4-3), as this redshift is seemingly consistent with the second peak of the NIR PDF(z). However, we found $z=3.66$ is very unlikely, because (1) the $z=3.66$ [CI](1-0) is not detected at the expected frequency 105.6~GHz ($<1\sigma$); (2) the SED fitting at $z=3.66$ yields an abnormally high dust mass to stellar mass fraction $0.09<{M_{\rm dust}/M_\ast}<0.13$ that is $>10\times$ above typical values \citep[e.g.,][]{Donevski2020}, again disfavouring the $z=3.66$ solution. Therefore, the multitude of evidence confirms the redshift of XS55 to be $z=4.8214\pm0.0004$.


\subsection{Morphology}
\label{sec:morphology}
We model the F277W and F444W morphology of XS55 using \texttt{Galfit} \citep{Peng2010_galfit}. We adopt three separate components in both F277W and F444W: a compact central emission, a diffuse central emission, and a companion component at the south-east. The NIRCam PSFs were obtained using Webbpsf \citep{Perrin2014_WebbPSF} with a pixel scale of $0.05{\rm \farcs/pix}$. We consider two cases; one case using a point source (compact component) together with two Sérsic profiles (diffuse component and companion), and another case with three separate Sérsic profiles. In the case using a point source and two Sérsic models in F444W, there is a ring of emission left in the residual (\cref{fig:morph}-middle-right), suggesting the brightest part of the galaxy is marginally resolved, or the PSF modelling is imperfect. In the case of three Sérsic profiles, there are no clear structures, as would be expected with random noise (\cref{fig:morph}-top). In the three Sérsic profile model of F444W, the compact component has a size of $0\farcs061\pm0\farcs007$ corresponding to $R_{\rm e}=0.40\pm0.05\,{\rm kpc}$ with Sérsic index $n<1.2$, while the diffuse emission and the companion have sizes and Sérsic indices of $R_{\rm e}=1.7\pm0.1\,{\rm kpc}$, $n=0.37\pm0.08$ and $R_{\rm e}=1.1\pm0.1\,{\rm kpc}$, $n=0.61\pm0.33$ respectively (\cref{tab:sizes}). In F444W, $33\pm3$\% of the total flux is coming from the compact component, while $53\pm3$\% is coming from the extended diffuse component, and $14\pm3$\% is coming from the companion.
In JWST/F277W, the fit of the compact component yields a bulge-like $n=2.6\pm0.8$ model with $R_{\rm e}=0.58\pm0.07\,{\rm kpc}$ (\cref{tab:sizes}), providing $87\pm2\%$ of the total flux. The diffuse component cannot be fit, and the fit to the companion yields a $n=2.0\pm1.4$ model with $R_{\rm e}=1.7\pm0.5\,{\rm kpc}$, providing $13\pm2\%$ of the total flux (\cref{sec:morphology,tab:sizes}). At this redshift, [{O{\tiny III}}]${\lambda\lambda4959,5007}$ and ${\rm H\beta}$ fall within the JWST/F277W filter, consequently the flux is possibly boosted by line emission (\cref{fig:full_sed}).  


\subsection{FIR SED}
To estimate the dust mass ($M_{\rm dust}$) and temperature ($T_{\rm dust}$), we fit the FIR and (sub)mm photometry of XS55 with a modified black-body (MBB) model \citep{Magdis2012SED} using the code \texttt{mercurius} \citep{Witstok2022-mercurius}. We explore two cases; one assuming optically thin dust, and the other assuming `self-consistent' optically thick dust. For the optically thick case, we place an upper limit on the emitting area based on the half light radius of the galaxy in JWST/F444W ($9.08\,{\rm kpc^2}$; \cref{sec:morphology}). The resulting fit and corresponding parameters are shown in \cref{fig:FIR-SEDs,tab:physpars}. In both cases, the IR spectral index $\beta_{\rm IR}$ is consistent with a weighted average of $\beta_{\rm IR}=2.02\,{\pm \,0.12}$. For the optically thin case, we recover a low dust temperature of $T_{\rm dust}=27.6^{+3.1}_{-2.7}\,{\rm K}$, with an accompanying high dust mass of $M_{\rm dust}=2.4^{+1.4}_{-0.9}\times10^9\,{\rm M_\odot}$. On the other hand, the optically thick dust yields a higher $T_{\rm dust}=32.5^{+2.2}_{-1.9}\,{\rm K}$ and a lower $M_{\rm dust}=1.7^{+0.5}_{-0.4}\times10^9\,{\rm M_\odot}$. 

\begin{figure*}[!htbp]
    \centering
    \includegraphics[width=0.8\textwidth]{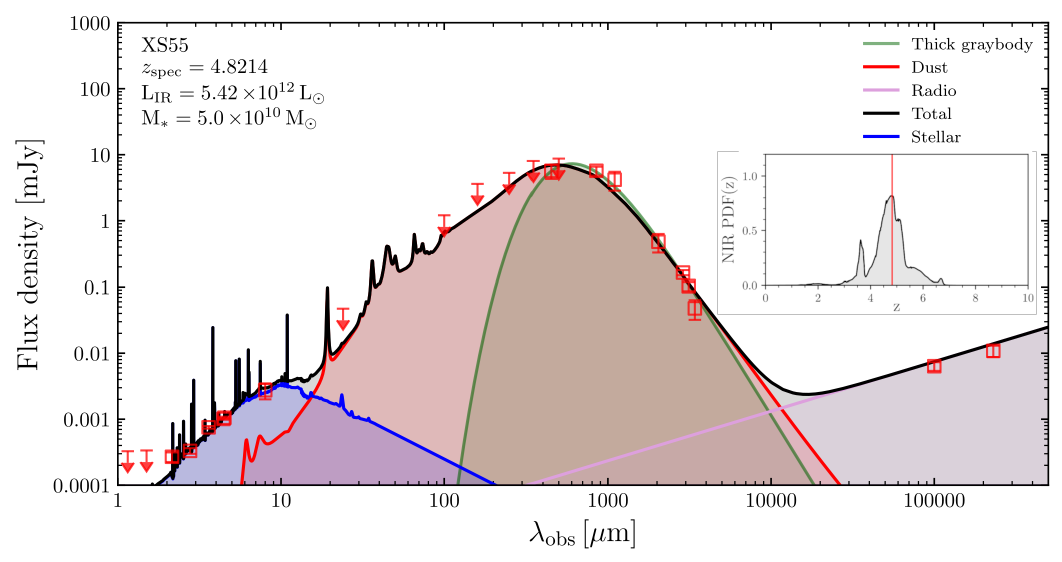}
    \caption{NIR to Radio SED of XS55, fit using \texttt{STARDUST} \citep{Kokorev2021}. The total SED (black), is shown with its different components: stellar (blue), dust (red), and radio (magenta). 
    An optically thick modified blackbody \citep{Magdis2012SED}, fitted with \texttt{mercurius} \citep{Witstok2022-mercurius}, is shown with a green line. The modified blackbody is not accounted for in the total SED fit.
    The radio component is extrapolated using the stellar mass dependent IR-Radio relation from \citet{Delvecchio2021_IR_Radio}. The NIR PDF(z) of XS55 is shown as an inset with the spec-z marked by a vertical red line.}
    \label{fig:full_sed}
\end{figure*}

To construct a panchromatic SED of XS55, we fit the stellar, AGN, and dust components using \texttt{STARDUST}
\citep{{Kokorev2021}} and available photometry from optical to radio wavelength, yielding an IR luminosity of $L_{\rm IR}=(5.43\pm1.77)\times10^{12}\,{\rm L_\odot}$ and a ${\rm SFR}_{\rm IR}=543\pm177\,{\rm M_\odot\,yr^{-1}}$. For the stellar component, we recover the stellar mass $\log(M_\ast/{\rm M_\odot})=10.7\pm0.1$, attenuated by $A_v=2.2\pm0.3\,{\rm mag}$. As an additional check, we also fit the stellar part of the spectrum (up to F444W) using  \texttt{Bagpipes}\footnote{\url{https://bagpipes.readthedocs.io/en/latest/}} \citep{Carnall2018_bagpipes}, using the same parameters as in \citet{Jin2024_cosbo7}, yielding physical properties similar to those from \texttt{STARDUST}, $\log(M_{\rm \ast,bagpipes}/{\rm M_\odot})=10.7\pm0.1$, $A_{\rm v,bagpipes}=2.1\pm0.3$. The physical parameters calculated with LePhare in the COSMOS-Web catalog (Shuntov et al. in prep.) also agree with our results ($\log(M_{\rm \ast,LePhare}/{\rm M_\odot})=10.8\pm0.3$, $E({\rm B-V})_{\rm LePhare}=1.1$).
The resulting panchromatic SED and corresponding properties from \texttt{STARDUST}, and MBB fit from \texttt{mercurius}, are shown in \cref{fig:full_sed,tab:physpars}.

\subsection{Molecular gas mass and far-infrared lines}
\label{sec:fir_lines}
With the derived dust mass, we infer a molecular gas mass ($M_{\rm mol}$) using the standard gas-to-dust mass ratio of star-forming galaxies \citep[$\delta_{\rm gdr}=100$, assuming solar metallicity,][]{Magdis2012SED}, yielding $M_{\rm mol,thin}=2.6^{+0.5}_{-0.4}\times10^{11}\,{\rm M_\odot}$ and $M_{\rm mol,thick}=1.7^{+0.5}_{-0.4}\times10^{11}\,{\rm M_\odot}$. Furthermore, based on the detection of the [C{\tiny I}](1-0) line (\cref{tab:measured_properties}), we infer the molecular gas mass $M_{\rm mol}/{\alpha_{\rm CI}}=(1.8\pm0.5)\times10^{11}/{\alpha_{\rm CI}}\,{\rm M_\odot}$, where $\alpha_{\rm C{\tiny I}}=17.0\pm0.3\,{\rm M_\odot\,K^{-1}\,km^{-1}\,s\,pc^{-2}}$ \citep{Dunne2022alphaCI} that is consistent with other calibrations \citep{Valentino2018CI,Heintz2020_alphaCI}. Adopting instead $\alpha_{\rm CI}=4.1\pm1.4\,{\rm M_\odot\,K^{-1}\,km^{-1}\,s\,pc^{-2}}$ from \citet{FriasCastillo2024} for high-$z$ SMGs, it yields a $M_{\rm mol}=(4.3\pm1.9)\times10^{10}\,{\rm M_\odot}$. 
Interestingly, the CO(5-4) emission is marginally resolved along the NE-SW direction (see \cref{fig:spectrum,fig:color-img}), as the minor axis of the CO emission as well as the continuum source are unresolved, we obtain upper limits on their sizes using Eq. (2) of \citet{Gomez_Guijarro2022}. We measure the size of the integrated CO(5-4) emission by fitting an elliptical Gaussian in $uv$-space with \texttt{GILDAS/uvfit}, and obtain $R_e=1.79\pm1.22\,{\rm kpc}$ with ${\rm PA}=28\pm20\,{\rm deg}$. We constructed the CO(5-4) moment-1 map using Cube Analysis and Rendering Tool for Astronomy \citep[\texttt{CARTA};][\cref{fig:color-img}-right]{Comrie2021_CARTA}.
In the moment-1 map there is a clear velocity gradient, with the highest redshifted velocity of $70\,{\rm km/s}$ and the highest blueshifted velocity of $-130\,{\rm km/s}$. This velocity gradient is comparable to that of the radio selected NIR-dark galaxies from \cite{Gentile2024}.
We discuss the possible interpretations of the extended CO(5-4) emission in Sect. 4.2.

\subsection{Obscured AGN}

\begin{figure*}[!htbp]
    \centering
    \includegraphics[width=\textwidth]{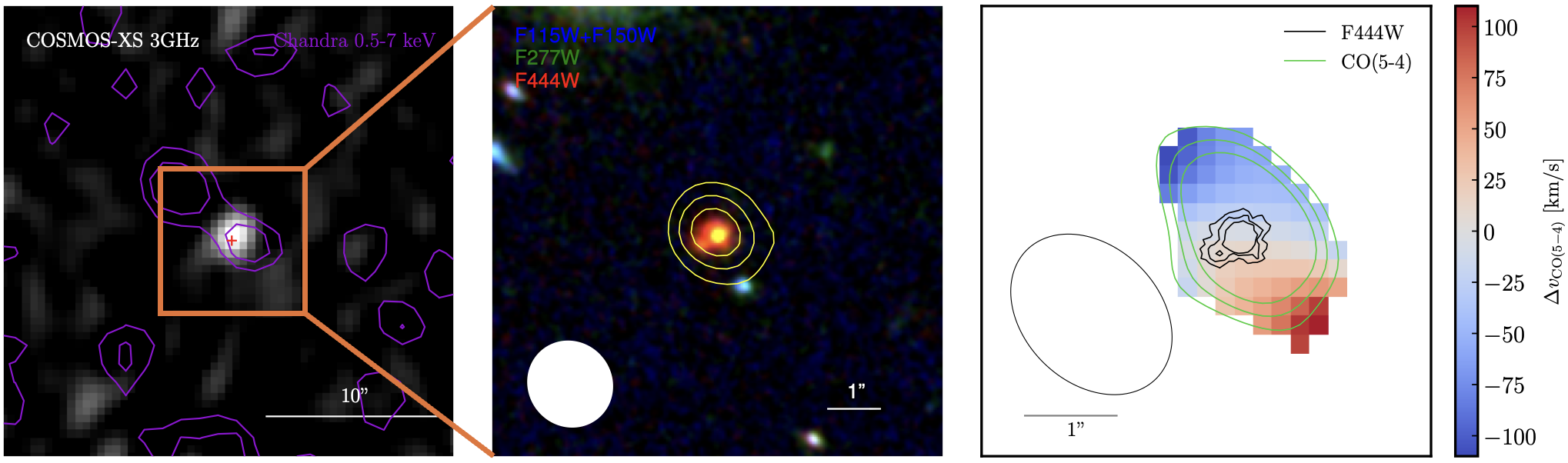}
    \caption{Multi-wavelength images of XS55.
    {\bf Left:} COSMOS-XS 3~GHz map \citep{van_der_Vlugt2021_COSMOS_XS} overlaid with 2,3$\sigma$ $0.5-7\,{\rm keV}$ contours from Chandra, smoothed with a 1" Gaussian, in purple. {\bf Middle:} JWST colour image of XS55 representing F115W+F105W, F277W, and F444W as blue, green, and red channels respectively. Overlaid are ALMA 3mm continuum emission contours at 5, 8, and 11$\sigma$, with the beam size shown as a white ellipse. {\bf Right:} moment-1 map of CO(5-4) masked at $3\sigma$ in moment-0, overlaid with integrated CO(5-4) at 3, 4, and 5$\sigma$ levels, and JWST/F444W contours at 5, 8, and 11$\sigma$ levels in yellow and black respectively.}
    \label{fig:color-img}
\end{figure*}

XS55 is tentatively detected in X-ray with both Chandra and XMM-Newton (Fig.~\ref{fig:color-img}, \ref{fig:cutouts}).
Using a 3" radius aperture that corresponds to the mean PSF of the Chandra COSMOS Legacy Survey \citep{Civano2016}, we detect 3 counts in the 0.5-2 keV band of Chandra. This corresponds to a flux of $f_{\rm [0.5-2keV]}=(1.15\pm0.54)\times10^{-16}\,{\rm erg\,s^{-1}\,cm^{-2}}$ adopting the conversion rate from \citet{Civano2016}. 
Since XS55 is extremely dusty, this flux should be considered a lower limit, yielding a soft X-ray luminosity of $L_{\rm [0.5-2keV]}>2.81\times10^{43}\,{\rm erg\,s^{-1}}$. Using the soft X-ray to bolometric luminosity correction from \citet{Lusso2012_Lbol} we obtained a bolometric luminosity of $L_{\rm bol}>5.35\times10^{44}\,{\rm erg\,s^{-1}}$. 
XS55 would be an X-ray selected AGN using the criterion from \citet{Riccio2023Xray} (i.e., $L_{0.2-2.3{\rm keV}}\geq3\times10^{42}\,{\rm erg\,s^{-1}}$). Due to the low significance of the X-ray detection, we cannot exclude that the detection could be spurious, or is originating from star-formation.
XS55 does not show excess radio emission compared to the infrared radio correlation (IRRC) of \citet{Delvecchio2021_IR_Radio}, however, this could be due to the large scatter of IR-radio correlation that is largely unconstrained at $z\sim5$ \citep{Delvecchio2021_IR_Radio}. Further, while radio excess is a clear indicator of AGN activity, X-ray AGN are not necessarily radio loud.



\begin{table}[!htbp]
    \centering
    \caption{Measured properties of XS55 emission lines}
    \setlength{\tabcolsep}{1.4pt}
    \renewcommand\arraystretch{1.5}
    \begin{tabular}{c c c c c c}
    \hline\hline
        Line & S/N &  FWHM  & $I_{\rm line}$ & $L'_{\rm line}$ & $P_{\rm chance}^a$ \\
        
        & & [${\rm km\,s^{-1}}$] & [${\rm Jy\,km\,s^{-1}}$]  & [$10^{10}{\rm K\,km\,s^{-1}\,pc^2}$] &  \\ \hline
        CO(5-4) & 9.8 & $365\pm37$ & $0.53\pm0.07$ & $1.9\pm0.2$ & $<10^{-6}$ \\
        ${\rm [C{\tiny I}]}$(1-0) & 3.9 & $293\pm75$ & $0.22\pm0.06$ & $1.1\pm0.3$ & $<0.004$ \\ \hline\hline
    \end{tabular}
    {\\Notes:$^a$chance probability of emission line \citep{Jin2019alma}.}
    \label{tab:measured_properties}
\end{table}

\begin{table}[!htbp]
    \caption{Fitted and inferred physical properties of XS55}
    \centering
    \setlength{\tabcolsep}{2pt}
    \renewcommand\arraystretch{1.5}
    \begin{tabular}{c c c c c c c c}
    \hline\hline
        Parameter & Value\\
        \hline
        ID  & XS55 \\
        \hline
        RA, Dec [deg]  & 150.1002501, 2.4967382 \\
        $z$ & $4.8214\pm0.0004$\\
        $A_V$ [mag] &  $2.2\pm0.3$ \\
        $M_\ast$ [$10^{10}{\rm M_\odot}$] & $5\pm1$\\
        ${\rm SFR}_{\rm IR}$ [${\rm M_\odot\,yr^{-1}}$] & $540\pm180$\\
        $L_{\rm X}$ [$10^{44}{\rm erg\,s^{-1}}$] & $>5.4$\\
        $L_{\rm IR}$ [$10^{12}{\rm L_{\odot}}$] & $5.4\pm1.8$\\
        $\beta_{\rm IR}$ & $2.0\pm0.2$\\
        $M_{\rm gas,[C{\tiny I}]}$ [$10^{11}{\rm M_\odot}$] & $1.8\pm0.5$\\
        ${\rm SFE}$ [$10^{-9}{\rm yr^{-1}}$] & $3.0\pm1.0$\\
        \hline
        $T_{\rm dust,thick}$ [K] & $32.7^{+2.2}_{-1.9}$\\
        $M_{\rm dust,thick}$ [$10^{9}{\rm M_\odot}$] & $1.7^{+0.6}_{-0.5}$ \\
        \hline
        $T_{\rm dust,thin}$ [K] & $28.0^{+3.3}_{-2.8}$\\
        $M_{\rm dust,thin}$ [$10^{9}{\rm M_\odot}$] & $2.3^{+1.4}_{-0.8}$ \\
        \hline
        $R_{\rm eff, 3mm}$ [kpc] & <1.76*\\
        $R_{\rm maj, CO(5-4))}$ [kpc] & $3.2\pm1.1$\\
        $R_{\rm min, CO(5-4))}$ [kpc] & <2.08*\\
        ${\rm PA}_{\rm CO(5-4)}$ [deg] & $28\pm20$\\
        
        \hline\hline
    \end{tabular}
    {\\Notes:*size upper limit (2$\sigma$) calculated with Eq. 2 in \citet{Gomez_Guijarro2022}.}
    \label{tab:physpars}
\end{table}

\section{Discussion}
\subsection{Intrinsically cold or optically thick dust (or both)?}
\begin{figure*}
    \centering
    \includegraphics[width=0.471\linewidth]{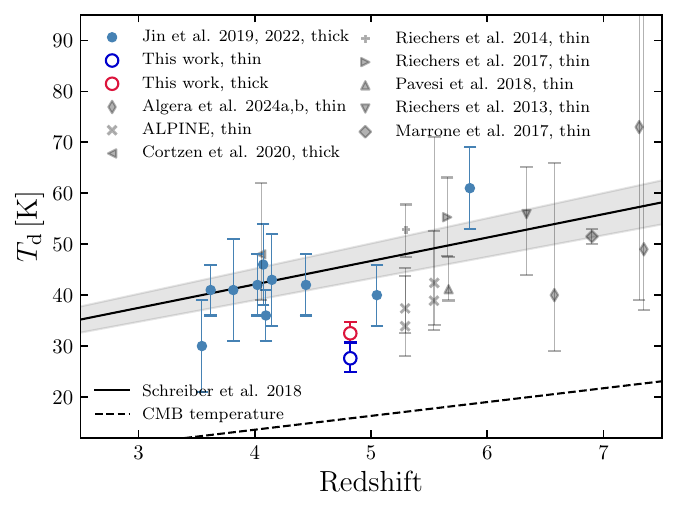}
    \includegraphics[width=0.48\textwidth]{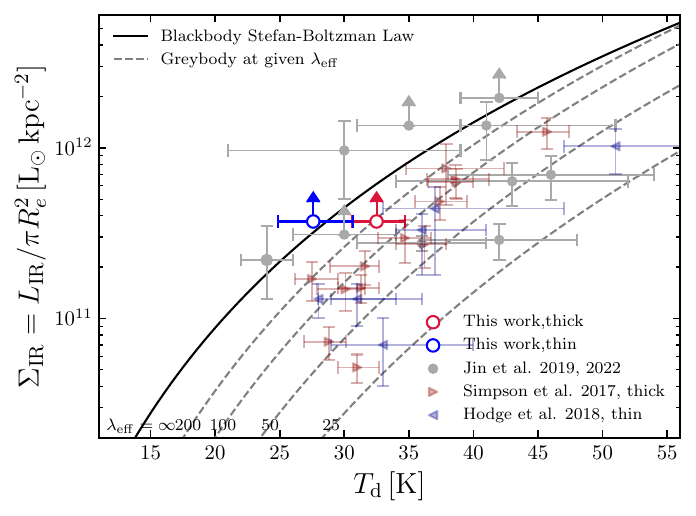}
    \caption{{\bf Left:} Dust temperature versus redshift for XS55 and literature samples. The $T_{\rm d}-z$ relation of main sequence galaxies from \citet{Schreiber2018Tdust} is shown as a black line, with the uncertainty as the gray shaded area. Literature samples are from \citet{Riechers2013Nature}, \citet{Riechers2014}, \citet{Riechers2017}, \citet{Marrone2017Nature},
    \citet{Pavesi2018}, \citet{Faisst2020}, \citet{Jin2019alma,Jin2022}, \citet{Fudamoto2023}, and \citet{Algera2024,Algera2024_tdust_rebels}. The CMB temperature as a function of redshift is shown as a dashed black line. {\bf Right:} Infrared luminosity surface density as a function of dust temperature, assuming optically thick dust. The surface density limit as defined by the Stefan-Boltzman law of optically thin dust is shown as a black line, while greybodies with varying $\lambda_{\rm eff}$ are shown as dashed lines. Literature samples are from \citet{Simpson2017}, \citet{Hodge2018}, and \citet{Jin2019alma,Jin2022}.} 
    \label{fig:Tdust_z_sigma}
\end{figure*}

To determine whether XS55 is intrinsically cold, or optically thick dust is making it appear cold, we apply three methods of diagnosing optically thick dust from \citet{Jin2022}:
(1) We compare the molecular gas masses derived from [C{\tiny I}](1-0) emission with gas mass inferred by both thin and thick $M_{\rm dust}$. We find that the lower gas mass from the thick dust model is in good agreement with the [C{\tiny I}]-derived gas mass, assuming $\delta_{\rm gdr}=100$ and $\alpha_{\rm [CI]}=17\pm3\,{\rm M_\odot\,K^{-1}\,km^{-1}\,s\,pc^{-2}}$, while the gas mass from the thin dust model is likely overestimated. 
(2) We estimate the dust opacity at 100\,$\mu$m, $\tau_{\rm 100\mu m}=\kappa \rho R_{\rm e}$, where $\kappa$ is adopted from \citet{Jones2013}, and $\rho$ is the volumetric dust density assuming spherical symmetry, using $R_{\rm e}$ from dust continuum ($R_{\rm e}<1.76\,{\rm kpc}$). This yields $\tau_{\rm 100\mu m,thin}>1.6$, and $\tau_{\rm 100\mu m,thick}>1.2$. Furthermore, we calculate the surface SFR density and find $\Sigma_{\rm SFR}>37\,{\rm M_\odot\,yr^{-1}\,kpc^{-2}}$. With both $\tau_{\rm 100\mu m}>1$ and $\Sigma_{\rm SFR}>20\,{\rm M_\odot\,yr^{-1}\,kpc^{-2}}$, the dust is hence optically thick \citep{Jin2022}.
(3) We derive a lower limit of IR luminosity surface density $\Sigma_{\rm IR}>3.7\times10^{11}\,{\rm L_\odot\,kpc^{-2}}$, and place it on the $\Sigma_{\rm IR}-T_{\rm dust}$ diagram. As seen in \cref{fig:Tdust_z_sigma}-right, the thin case is clearly violating the blackbody Stefan-Boltzmann law, and the optically thick case is more favourable.
Therefore, these pieces of evidence together imply that the dust of XS55 is optically thick in FIR.

Interestingly, the recovered $T_{\rm dust}$ assuming optically thick dust remains cold.
Comparing to the redshift evolution of $T_{\rm dust}$ in main sequence galaxies \citep{Schreiber2018Tdust,Jin2022}, the thick dust solution is 0.13 dex ($\sim4\sigma$) below the relation (\cref{fig:Tdust_z_sigma}). This makes XS55 one of the coldest DSFGs at $z>4$ discovered to date \citep[e.g.,][]{Faisst2020,Jin2022,Algera2024}.
Since dust temperature is proportional to the radiation field \citep[i.e., $\langle U\rangle=(T_{\rm d}/18.9)^{6.04}$,][]{Magdis2012SED}, and the radiation field is proportional to the ratio of star formation efficiency to metallicity \citep[i.e., $\langle U\rangle\propto {\rm SFE}/Z$,][]{Magdis2012SED}, and given that XS55 has typical SFE of DSFGs, the intrinsically low dust temperature could suggest a high metallicity in XS55, where the cooling is more efficient.



\subsection{Stable disk, outflow, or merger?}

Interestingly, the CO(5-4) is more extended than the dust continuum size and shows a tentative velocity gradient. These properties could suggest a rotating molecular disk \citep[e.g.,][]{Rizzo2023_alpaka} with compact star-formation \citep[e.g.,][]{Cochrane2019_compact_starburst}. However, given the low spatial resolution of the CO data, high resolution data are needed to confirm whether it is a rotating disk (e.g., \citealt{Rowland2024z7}). As suggested by the complex JWST morphology, galaxy interaction or merger can also be accounted for the extended CO(5-4) emission. Furthermore, given the possible AGN nature of XS55 and that the CO(5-4) is only resolved in one direction, a molecular outflow driven by a central AGN \citep[e.g.,][]{Lutz2020_Molecular_outflow} is also a potential scenario. 
To disentangle the above scenarios, a high resolution [CII] follow-up would be ideal to reveal the kinematics of XS55.

\subsection{Why is this source optically faint?}
In nearly all terms, XS55 is a normal but massive main sequence star-forming galaxy at $z=4.8214$, within 2$\sigma$ of the \citet{Schreiber2015} main-sequence relation (${\rm \Delta MS} ({\rm SFR/SFR_{MS}})=2.4\pm0.9$). Its star formation efficiency, and thereby depletion time of $300\pm100\,{\rm Myr}$, is comparable to other optically faint galaxies at similar redshift \citep[e.g.,][]{Jin2019alma,Jin2022}. However, with the compact stellar ($R_e=1.72\pm0.13\,{\rm kpc}$) and dust continuum ($R_e<1.76\,{\rm kpc}$) sizes, XS55 falls 2.6$\times$ below the mass-size relation of main sequence galaxies \citep{Ward2024_Mass_size_relation}. The compact size, together with a massive amount of dust, could explain the optically faint nature of this source, in agreement with observations from \citet{Gomez-Guijarro2023} and simulations from \citet{Cochrane2024_compact_hstdark}.

\section{Conclusions}
By combining ALMA spectroscopy with JWST imaging and multi-wavelength ancillary data in the COSMOS field, we presented a detailed study of a newly discovered radio selected optically faint galaxy, XS55. ALMA detections of the CO(5-4) and [C{\tiny I}](1-0) lines places the source at $z=4.8214$, and reveals the presence of large amounts of  dust in the ISM ($M_{\rm dust}=(1.7\pm0.5)\times10^9\,{\rm M_\odot}$). XS55 is a compact massive main-sequence galaxy, with extremely cold dust temperature, and optically thick FIR emission. The F444W size of the source falls below the mass-size relation. The tentative X-ray emission and the compactness of the central component suggest the presence of an AGN.  
The compactness together with the massive amount of cold dust in XS55 naturally explain its optical faintness.

\begin{acknowledgements}
    We thank the anonymous referee for constructive comments, improving this manuscript.
    This paper makes use of the following ALMA data: ADS/JAO.ALMA\#2022.1.00884.S, 2022.1.00863.S and 2021.1.00225.S. ALMA is a partnership of ESO (representing its member states), NSF (USA) and NINS (Japan), together with NRC (Canada), MOST and ASIAA (Taiwan), and KASI (Republic of Korea), in cooperation with the Republic of Chile. The Joint ALMA Observatory is operated by ESO, AUI/NRAO and NAOJ. Some of the data products presented herein were retrieved from the Dawn JWST Archive (DJA). DJA is an initiative of the Cosmic Dawn Center (DAWN). The Cosmic Dawn Center (DAWN) is funded by the Danish National Research Foundation under grant DNRF140.
    SJ acknowledges financial support from the European Union's Horizon Europe research and innovation program under the Marie Sk\l{}odowska-Curie grant No. 101060888. 
    GEM and SJ acknowledge the Villum Fonden research grants 37440 and 13160. JH acknowledges support from the ERC Consolidator Grant 101088676 (``VOYAJ''). CCC acknowledges the support of the Taiwan National Science and Technology Council (111-2112M-001-045-MY3), as well as the Academia Sinica through the Career Development Award (AS-CDA-112-M02). KK acknowledges support from the Knut and Alice Wallenberg Foundation (KAW 2017.0292 and 2019.0443).
\end{acknowledgements}

\bibliographystyle{aa}
\bibliography{biblio}

\begin{thebibliography}{87}
\expandafter\ifx\csname natexlab\endcsname\relax\def\natexlab#1{#1}\fi

\bibitem[{{Alcalde Pampliega} {et~al.}(2019){Alcalde Pampliega}, {P{\'e}rez-Gonz{\'a}lez}, {Barro}, {Dom{\'\i}nguez S{\'a}nchez}, {Eliche-Moral}, {Cardiel}, {Hern{\'a}n-Caballero}, {Rodriguez-Mu{\~n}oz}, {S{\'a}nchez Bl{\'a}zquez}, \& {Esquej}}]{Pampliega2019}
{Alcalde Pampliega}, B., {P{\'e}rez-Gonz{\'a}lez}, P.~G., {Barro}, G., {et~al.} 2019, \apj, 876, 135

\bibitem[{{Algera} {et~al.}(2024{\natexlab{a}}){Algera}, {Inami}, {De Looze}, {Ferrara}, {Hirashita}, {Aravena}, {Bakx}, {Bouwens}, {Bowler}, {Da Cunha}, {Dayal}, {Fudamoto}, {Hodge}, {Hygate}, {van Leeuwen}, {Nanayakkara}, {Palla}, {Pallottini}, {Rowland}, {Smit}, {Sommovigo}, {Stefanon}, {Vijayan}, \& {van der Werf}}]{Algera2024}
{Algera}, H. S.~B., {Inami}, H., {De Looze}, I., {et~al.} 2024{\natexlab{a}}, \mnras, 533, 3098

\bibitem[{{Algera} {et~al.}(2024{\natexlab{b}}){Algera}, {Inami}, {Sommovigo}, {Fudamoto}, {Schneider}, {Graziani}, {Dayal}, {Bouwens}, {Aravena}, {da Cunha}, {Ferrara}, {Hygate}, {van Leeuwen}, {De Looze}, {Palla}, {Pallottini}, {Smit}, {Stefanon}, {Topping}, \& {van der Werf}}]{Algera2024_tdust_rebels}
{Algera}, H. S.~B., {Inami}, H., {Sommovigo}, L., {et~al.} 2024{\natexlab{b}}, \mnras, 527, 6867

\bibitem[{{Algera} {et~al.}(2020){Algera}, {van der Vlugt}, {Hodge}, {Smail}, {Novak}, {Radcliffe}, {Riechers}, {R{\"o}ttgering}, {Smol{\v{c}}i{\'c}}, \& {Walter}}]{Algera2020}
{Algera}, H.~S.~B., {van der Vlugt}, D., {Hodge}, J.~A., {et~al.} 2020, \apj, 903, 139

\bibitem[{{Barrufet} {et~al.}(2024){Barrufet}, {Oesch}, {Marques-Chaves}, {Arellano-Cordova}, {Baggen}, {Carnall}, {Cullen}, {Dunlop}, {Gottumukkala}, {Fudamoto}, {Illingworth}, {Magee}, {McLure}, {McLeod}, {Micha{\l}owski}, {Stefanon}, {van Dokkum}, \& {Weibel}}]{Barrufet2024_OFGs}
{Barrufet}, L., {Oesch}, P., {Marques-Chaves}, R., {et~al.} 2024, arXiv e-prints, arXiv:2404.08052

\bibitem[{{Barrufet} {et~al.}(2023){Barrufet}, {Oesch}, {Weibel}, {Brammer}, {Bezanson}, {Bouwens}, {Fudamoto}, {Gonzalez}, {Gottumukkala}, {Illingworth}, {Heintz}, {Holden}, {Labbe}, {Magee}, {Naidu}, {Nelson}, {Stefanon}, {Smit}, {van Dokkum}, {Weaver}, \& {Williams}}]{Barrufet2023}
{Barrufet}, L., {Oesch}, P.~A., {Weibel}, A., {et~al.} 2023, \mnras, 522, 449

\bibitem[{{Birkin} {et~al.}(2021){Birkin}, {Weiss}, {Wardlow}, {Smail}, {Swinbank}, {Dudzevi{\v{c}}i{\={u}}t{\.{e}}}, {An}, {Ao}, {Chapman}, {Chen}, {da Cunha}, {Dannerbauer}, {Gullberg}, {Hodge}, {Ikarashi}, {Ivison}, {Matsuda}, {Stach}, {Walter}, {Wang}, \& {van der Werf}}]{Birkin2021}
{Birkin}, J.~E., {Weiss}, A., {Wardlow}, J.~L., {et~al.} 2021, \mnras, 501, 3926

\bibitem[{{Carnall} {et~al.}(2018){Carnall}, {McLure}, {Dunlop}, \& {Dav{\'e}}}]{Carnall2018_bagpipes}
{Carnall}, A.~C., {McLure}, R.~J., {Dunlop}, J.~S., \& {Dav{\'e}}, R. 2018, \mnras, 480, 4379

\bibitem[{{Casey} {et~al.}(2023){Casey}, {Kartaltepe}, {Drakos}, {Franco}, {Harish}, {Paquereau}, {Ilbert}, {Rose}, {Cox}, {Nightingale}, {Robertson}, {Silverman}, {Koekemoer}, {Massey}, {McCracken}, {Rhodes}, {Akins}, {Allen}, {Amvrosiadis}, {Arango-Toro}, {Bagley}, {Bongiorno}, {Capak}, {Champagne}, {Chartab}, {Ch{\'a}vez Ortiz}, {Chworowsky}, {Cooke}, {Cooper}, {Darvish}, {Ding}, {Faisst}, {Finkelstein}, {Fujimoto}, {Gentile}, {Gillman}, {Gould}, {Gozaliasl}, {Hayward}, {He}, {Hemmati}, {Hirschmann}, {Jahnke}, {Jin}, {Khostovan}, {Kokorev}, {Lambrides}, {Laigle}, {Larson}, {Leung}, {Liu}, {Liaudat}, {Long}, {Magdis}, {Mahler}, {Mainieri}, {Manning}, {Maraston}, {Martin}, {McCleary}, {McKinney}, {McPartland}, {Mobasher}, {Pattnaik}, {Renzini}, {Rich}, {Sanders}, {Sattari}, {Scognamiglio}, {Scoville}, {Sheth}, {Shuntov}, {Sparre}, {Suzuki}, {Talia}, {Toft}, {Trakhtenbrot}, {Urry}, {Valentino}, {Vanderhoof}, {Vardoulaki}, {Weaver}, {Whitaker}, {Wilkins}, {Yang}, \& {Zavala}}]{Casey2023_COSMOS_WEB}
{Casey}, C.~M., {Kartaltepe}, J.~S., {Drakos}, N.~E., {et~al.} 2023, \apj, 954, 31

\bibitem[{{Casey} {et~al.}(2019){Casey}, {Zavala}, {Aravena}, {B{\'e}thermin}, {Caputi}, {Champagne}, {Clements}, {da Cunha}, {Drew}, {Finkelstein}, {Hayward}, {Kartaltepe}, {Knudsen}, {Koekemoer}, {Magdis}, {Man}, {Manning}, {Scoville}, {Sheth}, {Spilker}, {Staguhn}, {Talia}, {Taniguchi}, {Toft}, {Treister}, \& {Yun}}]{Casey2019}
{Casey}, C.~M., {Zavala}, J.~A., {Aravena}, M., {et~al.} 2019, \apj, 887, 55

\bibitem[{{Casey} {et~al.}(2021){Casey}, {Zavala}, {Manning}, {Aravena}, {B{\'e}thermin}, {Caputi}, {Champagne}, {Clements}, {Drew}, {Finkelstein}, {Fujimoto}, {Hayward}, {Dekel}, {Kokorev}, {Lagos}, {Long}, {Magdis}, {Man}, {Mitsuhashi}, {Popping}, {Spilker}, {Staguhn}, {Talia}, {Toft}, {Treister}, {Weaver}, \& {Yun}}]{Casey2021_MORA}
{Casey}, C.~M., {Zavala}, J.~A., {Manning}, S.~M., {et~al.} 2021, \apj, 923, 215

\bibitem[{{Chabrier}(2003)}]{Chabrier2003}
{Chabrier}, G. 2003, \pasp, 115, 763

\bibitem[{{Chen} {et~al.}(2022){Chen}, {Liao}, {Smail}, {Swinbank}, {Ao}, {Bunker}, {Chapman}, {Hatsukade}, {Ivison}, {Lee}, {Serjeant}, {Umehata}, {Wang}, \& {Zhao}}]{Chen2022}
{Chen}, C.-C., {Liao}, C.-L., {Smail}, I., {et~al.} 2022, \apj, 929, 159

\bibitem[{{Civano} {et~al.}(2016){Civano}, {Marchesi}, {Comastri}, {Urry}, {Elvis}, {Cappelluti}, {Puccetti}, {Brusa}, {Zamorani}, {Hasinger}, {Aldcroft}, {Alexander}, {Allevato}, {Brunner}, {Capak}, {Finoguenov}, {Fiore}, {Fruscione}, {Gilli}, {Glotfelty}, {Griffiths}, {Hao}, {Harrison}, {Jahnke}, {Kartaltepe}, {Karim}, {LaMassa}, {Lanzuisi}, {Miyaji}, {Ranalli}, {Salvato}, {Sargent}, {Scoville}, {Schawinski}, {Schinnerer}, {Silverman}, {Smolcic}, {Stern}, {Toft}, {Trakhtenbrot}, {Treister}, \& {Vignali}}]{Civano2016}
{Civano}, F., {Marchesi}, S., {Comastri}, A., {et~al.} 2016, \apj, 819, 62

\bibitem[{{Cochrane} {et~al.}(2024){Cochrane}, {Angl{\'e}s-Alc{\'a}zar}, {Cullen}, \& {Hayward}}]{Cochrane2024_compact_hstdark}
{Cochrane}, R.~K., {Angl{\'e}s-Alc{\'a}zar}, D., {Cullen}, F., \& {Hayward}, C.~C. 2024, \apj, 961, 37

\bibitem[{{Cochrane} {et~al.}(2019){Cochrane}, {Hayward}, {Angl{\'e}s-Alc{\'a}zar}, {Lotz}, {Parsotan}, {Ma}, {Kere{\v{s}}}, {Feldmann}, {Faucher-Gigu{\`e}re}, \& {Hopkins}}]{Cochrane2019_compact_starburst}
{Cochrane}, R.~K., {Hayward}, C.~C., {Angl{\'e}s-Alc{\'a}zar}, D., {et~al.} 2019, \mnras, 488, 1779

\bibitem[{{Comrie} {et~al.}(2021){Comrie}, {Wang}, {Hsu}, {Moraghan}, {Harris}, {Pang}, {Pi{\'n}ska}, {Chiang}, {Chang}, {Hwang}, {Jan}, {Lin}, \& {Simmonds}}]{Comrie2021_CARTA}
{Comrie}, A., {Wang}, K.-S., {Hsu}, S.-C., {et~al.} 2021, {CARTA: The Cube Analysis and Rendering Tool for Astronomy}

\bibitem[{{Delvecchio} {et~al.}(2021){Delvecchio}, {Daddi}, {Sargent}, {Jarvis}, {Elbaz}, {Jin}, {Liu}, {Whittam}, {Algera}, {Carraro}, {D'Eugenio}, {Delhaize}, {Kalita}, {Leslie}, {Moln{\'a}r}, {Novak}, {Prandoni}, {Smol{\v{c}}i{\'c}}, {Ao}, {Aravena}, {Bournaud}, {Collier}, {Randriamampandry}, {Randriamanakoto}, {Rodighiero}, {Schober}, {White}, \& {Zamorani}}]{Delvecchio2021_IR_Radio}
{Delvecchio}, I., {Daddi}, E., {Sargent}, M.~T., {et~al.} 2021, \aap, 647, A123

\bibitem[{{Delvecchio} {et~al.}(2017){Delvecchio}, {Smol{\v c}i{\'c}}, {Zamorani}, {Lagos}, {Berta}, {Delhaize}, {Baran}, {Alexander}, {Rosario}, {Gonzalez-Perez}, {Ilbert}, {Lacey}, {Le F{\`e}vre}, {Miettinen}, {Aravena}, {Bondi}, {Carilli}, {Ciliegi}, {Mooley}, {Novak}, {Schinnerer}, {Capak}, {Civano}, {Fanidakis}, {Herrera Ruiz}, {Karim}, {Laigle}, {Marchesi}, {McCracken}, {Middleberg}, {Salvato}, \& {Tasca}}]{Delvecchio2017}
{Delvecchio}, I., {Smol{\v c}i{\'c}}, V., {Zamorani}, G., {et~al.} 2017, \aap, 602, A3

\bibitem[{{Donevski} {et~al.}(2020){Donevski}, {Lapi}, {Ma{\l}ek}, {Liu}, {G{\'o}mez-Guijarro}, {Dav{\'e}}, {Kraljic}, {Pantoni}, {Man}, {Fujimoto}, {Feltre}, {Pearson}, {Li}, \& {Narayanan}}]{Donevski2020}
{Donevski}, D., {Lapi}, A., {Ma{\l}ek}, K., {et~al.} 2020, \aap, 644, A144

\bibitem[{{Dunne} {et~al.}(2022){Dunne}, {Maddox}, {Papadopoulos}, {Ivison}, \& {Gomez}}]{Dunne2022alphaCI}
{Dunne}, L., {Maddox}, S.~J., {Papadopoulos}, P.~P., {Ivison}, R.~J., \& {Gomez}, H.~L. 2022, \mnras, 517, 962

\bibitem[{{Elvis} {et~al.}(2009){Elvis}, {Civano}, {Vignali}, {Puccetti}, {Fiore}, {Cappelluti}, {Aldcroft}, {Fruscione}, {Zamorani}, {Comastri}, {Brusa}, {Gilli}, {Miyaji}, {Damiani}, {Koekemoer}, {Finoguenov}, {Brunner}, {Urry}, {Silverman}, {Mainieri}, {Hasinger}, {Griffiths}, {Carollo}, {Hao}, {Guzzo}, {Blain}, {Calzetti}, {Carilli}, {Capak}, {Ettori}, {Fabbiano}, {Impey}, {Lilly}, {Mobasher}, {Rich}, {Salvato}, {Sanders}, {Schinnerer}, {Scoville}, {Shopbell}, {Taylor}, {Taniguchi}, \& {Volonteri}}]{Elvis2009_CCOSMOS}
{Elvis}, M., {Civano}, F., {Vignali}, C., {et~al.} 2009, \apjs, 184, 158

\bibitem[{{Enia} {et~al.}(2022){Enia}, {Talia}, {Pozzi}, {Cimatti}, {Delvecchio}, {Zamorani}, {D'Amato}, {Bisigello}, {Gruppioni}, {Rodighiero}, {Calura}, {Dallacasa}, {Giulietti}, {Barchiesi}, {Behiri}, \& {Romano}}]{Enia2022}
{Enia}, A., {Talia}, M., {Pozzi}, F., {et~al.} 2022, \apj, 927, 204

\bibitem[{{Faisst} {et~al.}(2020){Faisst}, {Fudamoto}, {Oesch}, {Scoville}, {Riechers}, {Pavesi}, \& {Capak}}]{Faisst2020}
{Faisst}, A.~L., {Fudamoto}, Y., {Oesch}, P.~A., {et~al.} 2020, \mnras, 498, 4192

\bibitem[{{Franco} {et~al.}(2018){Franco}, {Elbaz}, {B{\'e}thermin}, {Magnelli}, {Schreiber}, {Ciesla}, {Dickinson}, {Nagar}, {Silverman}, {Daddi}, {Alexander}, {Wang}, {Pannella}, {Le Floc'h}, {Pope}, {Giavalisco}, {Maury}, {Bournaud}, {Chary}, {Demarco}, {Ferguson}, {Finkelstein}, {Inami}, {Iono}, {Juneau}, {Lagache}, {Leiton}, {Lin}, {Magdis}, {Messias}, {Motohara}, {Mullaney}, {Okumura}, {Papovich}, {Pforr}, {Rujopakarn}, {Sargent}, {Shu}, \& {Zhou}}]{Franco2018}
{Franco}, M., {Elbaz}, D., {B{\'e}thermin}, M., {et~al.} 2018, \aap, 620, A152

\bibitem[{{Frias Castillo} {et~al.}(2024){Frias Castillo}, {Rybak}, {Hodge}, {Van der Werk}, {Smail}, {Butterworth}, {Jansen}, {Topkaras}, {Chen}, {Chapman}, {Weiss}, {Algera}, {Birkin}, {da Cunha}, {Chen}, {Dannerbauer}, {Jim{\'e}nez-Andrade}, {Ikarashi}, {Liao}, {Murphy}, {Swinbank}, {Walter}, {Calistro Rivera}, {Ivison}, \& {Lagos}}]{FriasCastillo2024}
{Frias Castillo}, M., {Rybak}, M., {Hodge}, J.~A., {et~al.} 2024, arXiv e-prints, arXiv:2404.05596

\bibitem[{{Fudamoto} {et~al.}(2023){Fudamoto}, {Inoue}, \& {Sugahara}}]{Fudamoto2023}
{Fudamoto}, Y., {Inoue}, A.~K., \& {Sugahara}, Y. 2023, \mnras, 521, 2962

\bibitem[{{Fudamoto} {et~al.}(2021){Fudamoto}, {Oesch}, {Schouws}, {Stefanon}, {Smit}, {Bouwens}, {Bowler}, {Endsley}, {Gonzalez}, {Inami}, {Labbe}, {Stark}, {Aravena}, {Barrufet}, {da Cunha}, {Dayal}, {Ferrara}, {Graziani}, {Hodge}, {Hutter}, {Li}, {De Looze}, {Nanayakkara}, {Pallottini}, {Riechers}, {Schneider}, {Ucci}, {van der Werf}, \& {White}}]{Fudamoto2021}
{Fudamoto}, Y., {Oesch}, P.~A., {Schouws}, S., {et~al.} 2021, \nat, 597, 489

\bibitem[{{Gao} {et~al.}(2024){Gao}, {Lim}, {Wang}, {Chen}, {Smail}, {Chapman}, {Zheng}, {Shim}, {Kodama}, {Ao}, {Chang}, {Clements}, {Dunlop}, {Ho}, {Hsu}, {Hwang}, {Hwang}, {Koprowski}, {Scott}, {Serjeant}, {Toba}, \& {Urquhart}}]{Gao2024_SCUBA450}
{Gao}, Z.-K., {Lim}, C.-F., {Wang}, W.-H., {et~al.} 2024, \apj, 971, 117

\bibitem[{{Gentile} {et~al.}(2024{\natexlab{a}}){Gentile}, {Talia}, {Behiri}, {Zamorani}, {Barchiesi}, {Vignali}, {Pozzi}, {Bethermin}, {Enia}, {Faisst}, {Giulietti}, {Gruppioni}, {Lapi}, {Massardi}, {Smol{\v{c}}i{\'c}}, {Vaccari}, \& {Cimatti}}]{Gentile2024a}
{Gentile}, F., {Talia}, M., {Behiri}, M., {et~al.} 2024{\natexlab{a}}, \apj, 962, 26

\bibitem[{{Gentile} {et~al.}(2024{\natexlab{b}}){Gentile}, {Talia}, {Daddi}, {Giulietti}, {Lapi}, {Massardi}, {Pozzi}, {Zamorani}, {Behiri}, {Enia}, {Bethermin}, {Dallacasa}, {Delvecchio}, {Faisst}, {Gruppioni}, {Loiacono}, {Traina}, {Vaccari}, {Vallini}, {Vignali}, {Smol{\v{c}}i{\'c}}, \& {Cimatti}}]{Gentile2024}
{Gentile}, F., {Talia}, M., {Daddi}, E., {et~al.} 2024{\natexlab{b}}, \aap, 687, A288

\bibitem[{{G{\'o}mez-Guijarro} {et~al.}(2022){G{\'o}mez-Guijarro}, {Elbaz}, {Xiao}, {B{\'e}thermin}, {Franco}, {Magnelli}, {Daddi}, {Dickinson}, {Demarco}, {Inami}, {Rujopakarn}, {Magdis}, {Shu}, {Chary}, {Zhou}, {Alexander}, {Bournaud}, {Ciesla}, {Ferguson}, {Finkelstein}, {Giavalisco}, {Iono}, {Juneau}, {Kartaltepe}, {Lagache}, {Le Floc'h}, {Leiton}, {Lin}, {Motohara}, {Mullaney}, {Okumura}, {Pannella}, {Papovich}, {Pope}, {Sargent}, {Silverman}, {Treister}, \& {Wang}}]{Gomez_Guijarro2022}
{G{\'o}mez-Guijarro}, C., {Elbaz}, D., {Xiao}, M., {et~al.} 2022, \aap, 658, A43

\bibitem[{{G{\'o}mez-Guijarro} {et~al.}(2023){G{\'o}mez-Guijarro}, {Magnelli}, {Elbaz}, {Wuyts}, {Daddi}, {Le Bail}, {Giavalisco}, {Dickinson}, {P{\'e}rez-Gonz{\'a}lez}, {Arrabal Haro}, {Bagley}, {Bisigello}, {Buat}, {Burgarella}, {Calabr{\`o}}, {Casey}, {Cheng}, {Ciesla}, {Dekel}, {Ferguson}, {Finkelstein}, {Franco}, {Grogin}, {Holwerda}, {Jin}, {Kartaltepe}, {Koekemoer}, {Kokorev}, {Long}, {Lucas}, {Magdis}, {Papovich}, {Pirzkal}, {Seill{\'e}}, {Tacchella}, {Tarrasse}, {Valentino}, {de la Vega}, {Wilkins}, {Xiao}, \& {Yung}}]{Gomez-Guijarro2023}
{G{\'o}mez-Guijarro}, C., {Magnelli}, B., {Elbaz}, D., {et~al.} 2023, \aap, 677, A34

\bibitem[{{Gottumukkala} {et~al.}(2024){Gottumukkala}, {Barrufet}, {Oesch}, {Weibel}, {Allen}, {Alcalde Pampliega}, {Nelson}, {Williams}, {Brammer}, {Fudamoto}, {Gonz{\'a}lez}, {Heintz}, {Illingworth}, {Magee}, {Naidu}, {Shuntov}, {Stefanon}, {Toft}, {Valentino}, \& {Xiao}}]{Gottumukkala2024_OFG_SMF}
{Gottumukkala}, R., {Barrufet}, L., {Oesch}, P.~A., {et~al.} 2024, \mnras, 530, 966

\bibitem[{{Hale} {et~al.}(2024){Hale}, {Heywood}, {Jarvis}, {Whittam}, {Best}, {An}, {Bowler}, {Harrison}, {Matthews}, {Smith}, {Taylor}, \& {Vaccari}}]{Hale2024_meerkat}
{Hale}, C.~L., {Heywood}, I., {Jarvis}, M.~J., {et~al.} 2024, \mnras [\eprint[arXiv]{2411.04958}]

\bibitem[{{Hasinger} {et~al.}(2007){Hasinger}, {Cappelluti}, {Brunner}, {Brusa}, {Comastri}, {Elvis}, {Finoguenov}, {Fiore}, {Franceschini}, {Gilli}, {Griffiths}, {Lehmann}, {Mainieri}, {Matt}, {Matute}, {Miyaji}, {Molendi}, {Paltani}, {Sanders}, {Scoville}, {Tresse}, {Urry}, {Vettolani}, \& {Zamorani}}]{Hasinger2007_XMMCosmos}
{Hasinger}, G., {Cappelluti}, N., {Brunner}, H., {et~al.} 2007, \apjs, 172, 29

\bibitem[{{Heintz} \& {Watson}(2020)}]{Heintz2020_alphaCI}
{Heintz}, K.~E. \& {Watson}, D. 2020, \apjl, 889, L7

\bibitem[{{Heywood} {et~al.}(2022){Heywood}, {Jarvis}, {Hale}, {Whittam}, {Bester}, {Hugo}, {Kenyon}, {Prescott}, {Smirnov}, {Tasse}, {Afonso}, {Best}, {Collier}, {Deane}, {Frank}, {Hardcastle}, {Knowles}, {Maddox}, {Murphy}, {Prandoni}, {Randriamampandry}, {Santos}, {Sekhar}, {Tabatabaei}, {Taylor}, \& {Thorat}}]{Heywood2022}
{Heywood}, I., {Jarvis}, M.~J., {Hale}, C.~L., {et~al.} 2022, \mnras, 509, 2150

\bibitem[{{Hodge} {et~al.}(2019){Hodge}, {Smail}, {Walter}, {da Cunha}, {Swinbank}, {Rybak}, {Venemans}, {Brandt}, {Calistro Rivera}, {Chapman}, {Chen}, {Cox}, {Dannerbauer}, {Decarli}, {Greve}, {Knudsen}, {Menten}, {Schinnerer}, {Simpson}, {van der Werf}, {Wardlow}, \& {Weiss}}]{Hodge2018}
{Hodge}, J.~A., {Smail}, I., {Walter}, F., {et~al.} 2019, \apj, 876, 130

\bibitem[{{Jarvis} {et~al.}(2016){Jarvis}, {Taylor}, {Agudo}, {Allison}, {Deane}, {Frank}, {Gupta}, {Heywood}, {Maddox}, {McAlpine}, {Santos}, {Scaife}, {Vaccari}, {Zwart}, {Adams}, {Bacon}, {Baker}, {Bassett}, {Best}, {Beswick}, {Blyth}, {Brown}, {Bruggen}, {Cluver}, {Colafrancesco}, {Cotter}, {Cress}, {Dav{\'e}}, {Ferrari}, {Hardcastle}, {Hale}, {Harrison}, {Hatfield}, {Klockner}, {Kolwa}, {Malefahlo}, {Marubini}, {Mauch}, {Moodley}, {Morganti}, {Norris}, {Peters}, {Prandoni}, {Prescott}, {Oliver}, {Oozeer}, {Rottgering}, {Seymour}, {Simpson}, {Smirnov}, \& {Smith}}]{Jarvis2016mightee}
{Jarvis}, M., {Taylor}, R., {Agudo}, I., {et~al.} 2016, in MeerKAT Science: On the Pathway to the SKA, 6

\bibitem[{{Jin} {et~al.}(2018){Jin}, {Daddi}, {Liu}, {Smol{\v c}i{\'c}}, {Schinnerer}, {Calabr{\`o}}, {Gu}, {Delhaize}, {Delvecchio}, {Gao}, {Salvato}, {Puglisi}, {Dickinson}, {Bertoldi}, {Sargent}, {Novak}, {Magdis}, {Aretxaga}, {Wilson}, \& {Capak}}]{Jin2018cosmos}
{Jin}, S., {Daddi}, E., {Liu}, D., {et~al.} 2018, \apj, 864, 56

\bibitem[{{Jin} {et~al.}(2019){Jin}, {Daddi}, {Magdis}, {Liu}, {Schinnerer}, {Papadopoulos}, {Gu}, {Gao}, \& {Calabr{\`o}}}]{Jin2019alma}
{Jin}, S., {Daddi}, E., {Magdis}, G.~E., {et~al.} 2019, \apj, 887, 144

\bibitem[{{Jin} {et~al.}(2022){Jin}, {Daddi}, {Magdis}, {Liu}, {Weaver}, {Tan}, {Valentino}, {Gao}, {Schinnerer}, {Calabr{\`o}}, {Gu}, \& {Sese}}]{Jin2022}
{Jin}, S., {Daddi}, E., {Magdis}, G.~E., {et~al.} 2022, \aap, 665, A3

\bibitem[{{Jin} {et~al.}(2024){Jin}, {Sillassen}, {Hodge}, {Magdis}, {Rizzo}, {Casey}, {Koekemoer}, {Valentino}, {Kokorev}, {Magnelli}, {Gobat}, {Gillman}, {Franco}, {Faisst}, {Kartaltepe}, {Schinnerer}, {Toft}, {Algera}, {Harish}, {Lee}, {Liu}, {Shuntov}, {Talia}, \& {Vijayan}}]{Jin2024_cosbo7}
{Jin}, S., {Sillassen}, N.~B., {Hodge}, J., {et~al.} 2024, \aap, 690, L16

\bibitem[{{Jones} {et~al.}(2013){Jones}, {Fanciullo}, {K{\"o}hler}, {Verstraete}, {Guillet}, {Bocchio}, \& {Ysard}}]{Jones2013}
{Jones}, A.~P., {Fanciullo}, L., {K{\"o}hler}, M., {et~al.} 2013, \aap, 558, A62

\bibitem[{{Kokorev} {et~al.}(2023){Kokorev}, {Jin}, {Magdis}, {Caputi}, {Valentino}, {Dayal}, {Trebitsch}, {Brammer}, {Fujimoto}, {Bauer}, {Iani}, {Kohno}, {Bl{\'a}nquez Ses{\'e}}, {G{\'o}mez-Guijarro}, {Rinaldi}, \& {Navarro-Carrera}}]{Kokorev2023_dustyguy}
{Kokorev}, V., {Jin}, S., {Magdis}, G.~E., {et~al.} 2023, \apjl, 945, L25

\bibitem[{{Kokorev} {et~al.}(2021){Kokorev}, {Magdis}, {Davidzon}, {Brammer}, {Valentino}, {Daddi}, {Ciesla}, {Liu}, {Jin}, {Cortzen}, {Delvecchio}, {Gim{\'e}nez-Arteaga}, {G{\'o}mez-Guijarro}, {Sargent}, {Toft}, \& {Weaver}}]{Kokorev2021}
{Kokorev}, V.~I., {Magdis}, G.~E., {Davidzon}, I., {et~al.} 2021, \apj, 921, 40

\bibitem[{{Ling} {et~al.}(2024){Ling}, {Sun}, {Cheng}, {Li}, {Ma}, \& {Yan}}]{Ling2024_cosbo7}
{Ling}, C., {Sun}, B., {Cheng}, C., {et~al.} 2024, \apjl, 969, L28

\bibitem[{{Liu} {et~al.}(2018){Liu}, {Daddi}, {Dickinson}, {Owen}, {Pannella}, {Sargent}, {B{\'e}thermin}, {Magdis}, {Gao}, {Shu}, {Wang}, {Jin}, \& {Inami}}]{Liu_DZ2017}
{Liu}, D., {Daddi}, E., {Dickinson}, M., {et~al.} 2018, \apj, 853, 172

\bibitem[{{Lusso} {et~al.}(2012){Lusso}, {Comastri}, {Simmons}, {Mignoli}, {Zamorani}, {Vignali}, {Brusa}, {Shankar}, {Lutz}, {Trump}, {Maiolino}, {Gilli}, {Bolzonella}, {Puccetti}, {Salvato}, {Impey}, {Civano}, {Elvis}, {Mainieri}, {Silverman}, {Koekemoer}, {Bongiorno}, {Merloni}, {Berta}, {Le Floc'h}, {Magnelli}, {Pozzi}, \& {Riguccini}}]{Lusso2012_Lbol}
{Lusso}, E., {Comastri}, A., {Simmons}, B.~D., {et~al.} 2012, \mnras, 425, 623

\bibitem[{{Lutz} {et~al.}(2020){Lutz}, {Sturm}, {Janssen}, {Veilleux}, {Aalto}, {Cicone}, {Contursi}, {Davies}, {Feruglio}, {Fischer}, {Fluetsch}, {Garcia-Burillo}, {Genzel}, {Gonz{\'a}lez-Alfonso}, {Graci{\'a}-Carpio}, {Herrera-Camus}, {Maiolino}, {Schruba}, {Shimizu}, {Sternberg}, {Tacconi}, \& {Wei{\ss}}}]{Lutz2020_Molecular_outflow}
{Lutz}, D., {Sturm}, E., {Janssen}, A., {et~al.} 2020, \aap, 633, A134

\bibitem[{{Magdis} {et~al.}(2012){Magdis}, {Daddi}, {B{\'e}thermin}, {Sargent}, {Elbaz}, {Pannella}, {Dickinson}, {Dannerbauer}, {da Cunha}, {Walter}, {Rigopoulou}, {Charmandaris}, {Hwang}, \& {Kartaltepe}}]{Magdis2012SED}
{Magdis}, G.~E., {Daddi}, E., {B{\'e}thermin}, M., {et~al.} 2012, \apj, 760, 6

\bibitem[{{Marrone} {et~al.}(2018){Marrone}, {Spilker}, {Hayward}, {Vieira}, {Aravena}, {Ashby}, {Bayliss}, {B{\'e}thermin}, {Brodwin}, {Bothwell}, {Carlstrom}, {Chapman}, {Chen}, {Crawford}, {Cunningham}, {De Breuck}, {Fassnacht}, {Gonzalez}, {Greve}, {Hezaveh}, {Lacaille}, {Litke}, {Lower}, {Ma}, {Malkan}, {Miller}, {Morningstar}, {Murphy}, {Narayanan}, {Phadke}, {Rotermund}, {Sreevani}, {Stalder}, {Stark}, {Strandet}, {Tang}, \& {Wei{\ss}}}]{Marrone2017Nature}
{Marrone}, D.~P., {Spilker}, J.~S., {Hayward}, C.~C., {et~al.} 2018, \nat, 553, 51

\bibitem[{{McMullin} {et~al.}(2007){McMullin}, {Waters}, {Schiebel}, {Young}, \& {Golap}}]{McMullin2007CASA}
{McMullin}, J.~P., {Waters}, B., {Schiebel}, D., {Young}, W., \& {Golap}, K. 2007, in Astronomical Society of the Pacific Conference Series, Vol. 376, Astronomical Data Analysis Software and Systems XVI, ed. R.~A. {Shaw}, F.~{Hill}, \& D.~J. {Bell}, 127

\bibitem[{{Oke}(1974)}]{Oke1974_AB_sys}
{Oke}, J.~B. 1974, \apjs, 27, 21

\bibitem[{{Pavesi} {et~al.}(2018){Pavesi}, {Riechers}, {Sharon}, {Smol{\v c}i{\'c}}, {Faisst}, {Schinnerer}, {Carilli}, {Capak}, {Scoville}, \& {Stacey}}]{Pavesi2018}
{Pavesi}, R., {Riechers}, D.~A., {Sharon}, C.~E., {et~al.} 2018, \apj, 861, 43

\bibitem[{{Peng} {et~al.}(2010){Peng}, {Ho}, {Impey}, \& {Rix}}]{Peng2010_galfit}
{Peng}, C.~Y., {Ho}, L.~C., {Impey}, C.~D., \& {Rix}, H.-W. 2010, \aj, 139, 2097

\bibitem[{{P{\'e}rez-Gonz{\'a}lez} {et~al.}(2023){P{\'e}rez-Gonz{\'a}lez}, {Barro}, {Annunziatella}, {Costantin}, {Garc{\'\i}a-Argum{\'a}nez}, {McGrath}, {M{\'e}rida}, {Zavala}, {Arrabal Haro}, {Bagley}, {Backhaus}, {Behroozi}, {Bell}, {Bisigello}, {Buat}, {Calabr{\`o}}, {Casey}, {Cleri}, {Coogan}, {Cooper}, {Cooray}, {Dekel}, {Dickinson}, {Elbaz}, {Ferguson}, {Finkelstein}, {Fontana}, {Franco}, {Gardner}, {Giavalisco}, {G{\'o}mez-Guijarro}, {Grazian}, {Grogin}, {Guo}, {Huertas-Company}, {Jogee}, {Kartaltepe}, {Kewley}, {Kirkpatrick}, {Kocevski}, {Koekemoer}, {Long}, {Lotz}, {Lucas}, {Papovich}, {Pirzkal}, {Ravindranath}, {Somerville}, {Tacchella}, {Trump}, {Wang}, {Wilkins}, {Wuyts}, {Yang}, \& {Yung}}]{Perez-Gonzalez2023}
{P{\'e}rez-Gonz{\'a}lez}, P.~G., {Barro}, G., {Annunziatella}, M., {et~al.} 2023, \apjl, 946, L16

\bibitem[{{Perrin} {et~al.}(2014){Perrin}, {Sivaramakrishnan}, {Lajoie}, {Elliott}, {Pueyo}, {Ravindranath}, \& {Albert}}]{Perrin2014_WebbPSF}
{Perrin}, M.~D., {Sivaramakrishnan}, A., {Lajoie}, C.-P., {et~al.} 2014, in Society of Photo-Optical Instrumentation Engineers (SPIE) Conference Series, Vol. 9143, Space Telescopes and Instrumentation 2014: Optical, Infrared, and Millimeter Wave, ed. J.~{Oschmann}, Jacobus~M., M.~{Clampin}, G.~G. {Fazio}, \& H.~A. {MacEwen}, 91433X

\bibitem[{{Riccio} {et~al.}(2023){Riccio}, {Yang}, {Ma{\l}ek}, {Boquien}, {Junais}, {Pistis}, {Hamed}, {Grespan}, {Paolillo}, \& {Torbaniuk}}]{Riccio2023Xray}
{Riccio}, G., {Yang}, G., {Ma{\l}ek}, K., {et~al.} 2023, \aap, 678, A164

\bibitem[{{Riechers} {et~al.}(2013){Riechers}, {Bradford}, {Clements}, {Dowell}, {P{\'e}rez-Fournon}, {Ivison}, {Bridge}, {Conley}, {Fu}, {Vieira}, {Wardlow}, {Calanog}, {Cooray}, {Hurley}, {Neri}, {Kamenetzky}, {Aguirre}, {Altieri}, {Arumugam}, {Benford}, {B{\'e}thermin}, {Bock}, {Burgarella}, {Cabrera-Lavers}, {Chapman}, {Cox}, {Dunlop}, {Earle}, {Farrah}, {Ferrero}, {Franceschini}, {Gavazzi}, {Glenn}, {Solares}, {Gurwell}, {Halpern}, {Hatziminaoglou}, {Hyde}, {Ibar}, {Kov{\'a}cs}, {Krips}, {Lupu}, {Maloney}, {Martinez-Navajas}, {Matsuhara}, {Murphy}, {Naylor}, {Nguyen}, {Oliver}, {Omont}, {Page}, {Petitpas}, {Rangwala}, {Roseboom}, {Scott}, {Smith}, {Staguhn}, {Streblyanska}, {Thomson}, {Valtchanov}, {Viero}, {Wang}, {Zemcov}, \& {Zmuidzinas}}]{Riechers2013Nature}
{Riechers}, D.~A., {Bradford}, C.~M., {Clements}, D.~L., {et~al.} 2013, \nat, 496, 329

\bibitem[{{Riechers} {et~al.}(2014){Riechers}, {Carilli}, {Capak}, {Scoville}, {Smol{\v c}i{\'c}}, {Schinnerer}, {Yun}, {Cox}, {Bertoldi}, {Karim}, \& {Yan}}]{Riechers2014}
{Riechers}, D.~A., {Carilli}, C.~L., {Capak}, P.~L., {et~al.} 2014, \apj, 796, 84

\bibitem[{{Riechers} {et~al.}(2017){Riechers}, {Leung}, {Ivison}, {P{\'e}rez-Fournon}, {Lewis}, {Marques-Chaves}, {Oteo}, {Clements}, {Cooray}, {Greenslade}, {Mart{\'{\i}}nez-Navajas}, {Oliver}, {Rigopoulou}, {Scott}, \& {Weiss}}]{Riechers2017}
{Riechers}, D.~A., {Leung}, T.~K.~D., {Ivison}, R.~J., {et~al.} 2017, \apj, 850, 1

\bibitem[{{Rizzo} {et~al.}(2023){Rizzo}, {Roman-Oliveira}, {Fraternali}, {Frickmann}, {Valentino}, {Brammer}, {Zanella}, {Kokorev}, {Popping}, {Whitaker}, {Kohandel}, {Magdis}, {Di Mascolo}, {Ikeda}, {Jin}, \& {Toft}}]{Rizzo2023_alpaka}
{Rizzo}, F., {Roman-Oliveira}, F., {Fraternali}, F., {et~al.} 2023, \aap, 679, A129

\bibitem[{{Rowland} {et~al.}(2024){Rowland}, {Hodge}, {Bouwens}, {Pi{\~n}a}, {Hygate}, {Algera}, {Aravena}, {Bowler}, {da Cunha}, {Dayal}, {Ferrara}, {Herard-Demanche}, {Inami}, {van Leeuwen}, {de Looze}, {Oesch}, {Pallottini}, {Phillips}, {Rybak}, {Schouws}, {Smit}, {Sommovigo}, {Stefanon}, \& {van der Werf}}]{Rowland2024z7}
{Rowland}, L.~E., {Hodge}, J., {Bouwens}, R., {et~al.} 2024, \mnras [\eprint[arXiv]{2405.06025}]

\bibitem[{{Schreiber} {et~al.}(2018){Schreiber}, {Elbaz}, {Pannella}, {Ciesla}, {Wang}, \& {Franco}}]{Schreiber2018Tdust}
{Schreiber}, C., {Elbaz}, D., {Pannella}, M., {et~al.} 2018, \aap, 609, A30

\bibitem[{{Schreiber} {et~al.}(2015){Schreiber}, {Pannella}, {Elbaz}, {B{\'e}thermin}, {Inami}, {Dickinson}, {Magnelli}, {Wang}, {Aussel}, {Daddi}, {Juneau}, {Shu}, {Sargent}, {Buat}, {Faber}, {Ferguson}, {Giavalisco}, {Koekemoer}, {Magdis}, {Morrison}, {Papovich}, {Santini}, \& {Scott}}]{Schreiber2015}
{Schreiber}, C., {Pannella}, M., {Elbaz}, D., {et~al.} 2015, \aap, 575, A74

\bibitem[{{Shu} {et~al.}(2022){Shu}, {Yang}, {Liu}, {Wang}, {Wang}, {Han}, {Huang}, {Lim}, {Chang}, {Zheng}, {Zheng}, {Wang}, \& {Kong}}]{Shu2022}
{Shu}, X., {Yang}, L., {Liu}, D., {et~al.} 2022, \apj, 926, 155

\bibitem[{{Sillassen} {et~al.}(2024){Sillassen}, {Jin}, {Magdis}, {Daddi}, {Wang}, {Lu}, {Sun}, {Arumugam}, {Liu}, {Brinch}, {D'Eugenio}, {Gobat}, {G{\'o}mez-Guijarro}, {Rich}, {Schinnerer}, {Strazzullo}, {Tan}, {Valentino}, {Wang}, {Xiao}, {Zhou}, {Bl{\'a}nquez-Ses{\'e}}, {Cai}, {Chen}, {Ciesla}, {Dai}, {Delvecchio}, {Elbaz}, {Finoguenov}, {Gao}, {Gu}, {Hale}, {Hao}, {Huang}, {Jarvis}, {Kalita}, {Ke}, {Le Bail}, {Magnelli}, {Shi}, {Vaccari}, {Whittam}, {Yang}, \& {Zhang}}]{Sillassen2024}
{Sillassen}, N.~B., {Jin}, S., {Magdis}, G.~E., {et~al.} 2024, \aap, 690, A55

\bibitem[{{Simpson} {et~al.}(2019){Simpson}, {Smail}, {Swinbank}, {Chapman}, {Chen}, {Geach}, {Matsuda}, {Wang}, {Wang}, {Yang}, {Ao}, {Asquith}, {Bourne}, {Coogan}, {Coppin}, {Gullberg}, {Hine}, {Ho}, {Hwang}, {Ivison}, {Kato}, {Lacaille}, {Lewis}, {Liu}, {Micha{\l}owski}, {Oteo}, {Sawicki}, {Scholtz}, {Smith}, {Thomson}, \& {Wardlow}}]{Simpson2019}
{Simpson}, J.~M., {Smail}, I., {Swinbank}, A.~M., {et~al.} 2019, \apj, 880, 43

\bibitem[{{Simpson} {et~al.}(2017){Simpson}, {Smail}, {Swinbank}, {Ivison}, {Dunlop}, {Geach}, {Almaini}, {Arumugam}, {Bremer}, {Chen}, {Conselice}, {Coppin}, {Farrah}, {Ibar}, {Hartley}, {Ma}, {Micha{\l}owski}, {Scott}, {Spaans}, {Thomson}, \& {van der Werf}}]{Simpson2017}
{Simpson}, J.~M., {Smail}, I., {Swinbank}, A.~M., {et~al.} 2017, \apj, 839, 58

\bibitem[{{Smail} {et~al.}(2023){Smail}, {Dudzevi{\v{c}}i{\={u}}t{\.{e}}}, {Gurwell}, {Fazio}, {Willner}, {Swinbank}, {Arumugam}, {Summers}, {Cohen}, {Jansen}, {Windhorst}, {Meena}, {Zitrin}, {Keel}, {Cheng}, {Coe}, {Conselice}, {D'Silva}, {Driver}, {Frye}, {Grogin}, {Koekemoer}, {Marshall}, {Nonino}, {Pirzkal}, {Robotham}, {Rutkowski}, {Ryan}, {Tompkins}, {Willmer}, {Yan}, {Broadhurst}, {Diego}, {Kamieneski}, \& {Yun}}]{Smail2023}
{Smail}, I., {Dudzevi{\v{c}}i{\={u}}t{\.{e}}}, U., {Gurwell}, M., {et~al.} 2023, \apj, 958, 36

\bibitem[{{Smail} {et~al.}(2021){Smail}, {Dudzevi{\v{c}}i{\={u}}t{\.{e}}}, {Stach}, {Almaini}, {Birkin}, {Chapman}, {Chen}, {Geach}, {Gullberg}, {Hodge}, {Ikarashi}, {Ivison}, {Scott}, {Simpson}, {Swinbank}, {Thomson}, {Walter}, {Wardlow}, \& {van der Werf}}]{Smail2020}
{Smail}, I., {Dudzevi{\v{c}}i{\={u}}t{\.{e}}}, U., {Stach}, S.~M., {et~al.} 2021, \mnras, 502, 3426

\bibitem[{{Smol{\v c}i{\'c}} {et~al.}(2017){Smol{\v c}i{\'c}}, {Novak}, {Bondi}, {Ciliegi}, {Mooley}, {Schinnerer}, {Zamorani}, {Navarrete}, {Bourke}, {Karim}, {Vardoulaki}, {Leslie}, {Delhaize}, {Carilli}, {Myers}, {Baran}, {Delvecchio}, {Miettinen}, {Banfield}, {Balokovi{\'c}}, {Bertoldi}, {Capak}, {Frail}, {Hallinan}, {Hao}, {Herrera Ruiz}, {Horesh}, {Ilbert}, {Intema}, {Jeli{\'c}}, {Kl{\"o}ckner}, {Krpan}, {Kulkarni}, {McCracken}, {Laigle}, {Middleberg}, {Murphy}, {Sargent}, {Scoville}, \& {Sheth}}]{Smolcic2017}
{Smol{\v c}i{\'c}}, V., {Novak}, M., {Bondi}, M., {et~al.} 2017, \aap, 602, A1

\bibitem[{{Talia} {et~al.}(2021){Talia}, {Cimatti}, {Giulietti}, {Zamorani}, {Bethermin}, {Faisst}, {Le F{\`e}vre}, \& {Smol{\c{c}}i{\'c}}}]{Talia2021}
{Talia}, M., {Cimatti}, A., {Giulietti}, M., {et~al.} 2021, \apj, 909, 23

\bibitem[{{Valentino} {et~al.}(2023){Valentino}, {Brammer}, {Gould}, {Kokorev}, {Fujimoto}, {Jespersen}, {Vijayan}, {Weaver}, {Ito}, {Tanaka}, {Ilbert}, {Magdis}, {Whitaker}, {Faisst}, {Gallazzi}, {Gillman}, {Gim{\'e}nez-Arteaga}, {G{\'o}mez-Guijarro}, {Kubo}, {Heintz}, {Hirschmann}, {Oesch}, {Onodera}, {Rizzo}, {Lee}, {Strait}, \& {Toft}}]{Valentino2023_DJA}
{Valentino}, F., {Brammer}, G., {Gould}, K. M.~L., {et~al.} 2023, \apj, 947, 20

\bibitem[{{Valentino} {et~al.}(2018){Valentino}, {Magdis}, {Daddi}, {Liu}, {Aravena}, {Bournaud}, {Cibinel}, {Cormier}, {Dickinson}, {Gao}, {Jin}, {Juneau}, {Kartaltepe}, {Lee}, {Madden}, {Puglisi}, {Sanders}, \& {Silverman}}]{Valentino2018CI}
{Valentino}, F., {Magdis}, G.~E., {Daddi}, E., {et~al.} 2018, \apj, 869, 27

\bibitem[{{van der Vlugt} {et~al.}(2021){van der Vlugt}, {Algera}, {Hodge}, {Novak}, {Radcliffe}, {Riechers}, {R{\"o}ttgering}, {Smol{\v{c}}i{\'c}}, \& {Walter}}]{van_der_Vlugt2021_COSMOS_XS}
{van der Vlugt}, D., {Algera}, H.~S.~B., {Hodge}, J.~A., {et~al.} 2021, \apj, 907, 5

\bibitem[{{van der Vlugt} {et~al.}(2023){van der Vlugt}, {Hodge}, {Jin}, {Algera}, {Leslie}, {Riechers}, {R{\"o}ttgering}, {Smol{\v{c}}i{\'c}}, \& {Walter}}]{vanderVlugt2023_COSMOS_XS}
{van der Vlugt}, D., {Hodge}, J.~A., {Jin}, S., {et~al.} 2023, \apj, 951, 131

\bibitem[{{Walter} {et~al.}(2012){Walter}, {Decarli}, {Carilli}, {Bertoldi}, {Cox}, {da Cunha}, {Daddi}, {Dickinson}, {Downes}, {Elbaz}, {Ellis}, {Hodge}, {Neri}, {Riechers}, {Weiss}, {Bell}, {Dannerbauer}, {Krips}, {Krumholz}, {Lentati}, {Maiolino}, {Menten}, {Rix}, {Robertson}, {Spinrad}, {Stark}, \& {Stern}}]{Walter2012}
{Walter}, F., {Decarli}, R., {Carilli}, C., {et~al.} 2012, \nat, 486, 233

\bibitem[{{Wang} {et~al.}(2019){Wang}, {Schreiber}, {Elbaz}, {Yoshimura}, {Kohno}, {Shu}, {Yamaguchi}, {Pannella}, {Franco}, {Huang}, {Lim}, \& {Wang}}]{Wang2019Natur}
{Wang}, T., {Schreiber}, C., {Elbaz}, D., {et~al.} 2019, \nat, 572, 211

\bibitem[{{Ward} {et~al.}(2024){Ward}, {de la Vega}, {Mobasher}, {McGrath}, {Iyer}, {Calabr{\`o}}, {Costantin}, {Dickinson}, {Holwerda}, {Huertas-Company}, {Hirschmann}, {Lucas}, {Pandya}, {Wilkins}, {Yung}, {Arrabal Haro}, {Bagley}, {Finkelstein}, {Kartaltepe}, {Koekemoer}, {Papovich}, \& {Pirzkal}}]{Ward2024_Mass_size_relation}
{Ward}, E., {de la Vega}, A., {Mobasher}, B., {et~al.} 2024, \apj, 962, 176

\bibitem[{{Weaver} {et~al.}(2022){Weaver}, {Kauffmann}, {Ilbert}, {McCracken}, {Moneti}, {Toft}, {Brammer}, {Shuntov}, {Davidzon}, {Hsieh}, {Laigle}, {Anastasiou}, {Jespersen}, {Vinther}, {Capak}, {Casey}, {McPartland}, {Milvang-Jensen}, {Mobasher}, {Sanders}, {Zalesky}, {Arnouts}, {Aussel}, {Dunlop}, {Faisst}, {Franx}, {Furtak}, {Fynbo}, {Gould}, {Greve}, {Gwyn}, {Kartaltepe}, {Kashino}, {Koekemoer}, {Kokorev}, {Le F{\`e}vre}, {Lilly}, {Masters}, {Magdis}, {Mehta}, {Peng}, {Riechers}, {Salvato}, {Sawicki}, {Scarlata}, {Scoville}, {Shirley}, {Silverman}, {Sneppen}, {Smolc̆i{\'c}}, {Steinhardt}, {Stern}, {Tanaka}, {Taniguchi}, {Teplitz}, {Vaccari}, {Wang}, \& {Zamorani}}]{Weaver2022COSMOS2020}
{Weaver}, J.~R., {Kauffmann}, O.~B., {Ilbert}, O., {et~al.} 2022, \apjs, 258, 11

\bibitem[{{Wei{\ss}} {et~al.}(2009){Wei{\ss}}, {Ivison}, {Downes}, {Walter}, {Cirasuolo}, \& {Menten}}]{Weiss2009}
{Wei{\ss}}, A., {Ivison}, R.~J., {Downes}, D., {et~al.} 2009, \apjl, 705, L45

\bibitem[{{Witstok} {et~al.}(2022){Witstok}, {Smit}, {Maiolino}, {Kumari}, {Aravena}, {Boogaard}, {Bouwens}, {Carniani}, {Hodge}, {Jones}, {Stefanon}, {van der Werf}, \& {Schouws}}]{Witstok2022-mercurius}
{Witstok}, J., {Smit}, R., {Maiolino}, R., {et~al.} 2022, \mnras, 515, 1751

\bibitem[{{Xiao} {et~al.}(2023){Xiao}, {Elbaz}, {G{\'o}mez-Guijarro}, {Leroy}, {Bing}, {Daddi}, {Magnelli}, {Franco}, {Zhou}, {Dickinson}, {Wang}, {Rujopakarn}, {Magdis}, {Treister}, {Inami}, {Demarco}, {Sargent}, {Shu}, {Kartaltepe}, {Alexander}, {B{\'e}thermin}, {Bournaud}, {Ciesla}, {Ferguson}, {Finkelstein}, {Giavalisco}, {Gu}, {Iono}, {Juneau}, {Lagache}, {Leiton}, {Messias}, {Motohara}, {Mullaney}, {Nagar}, {Pannella}, {Papovich}, {Pope}, {Schreiber}, \& {Silverman}}]{Xiao2023_OFGs}
{Xiao}, M.~Y., {Elbaz}, D., {G{\'o}mez-Guijarro}, C., {et~al.} 2023, \aap, 672, A18

\bibitem[{{Zhou} {et~al.}(2024){Zhou}, {Wang}, {Daddi}, {Coogan}, {Sun}, {Xu}, {Arumugam}, {Jin}, {Liu}, {Lu}, {Sillassen}, {Wang}, {Shi}, {Zhang}, {Tan}, {Gu}, {Elbaz}, {Le Bail}, {Magnelli}, {G{\'o}mez-Guijarro}, {d'Eugenio}, {Magdis}, {Valentino}, {Ji}, {Gobat}, {Delvecchio}, {Xiao}, {Strazzullo}, {Finoguenov}, {Schinnerer}, {Rich}, {Huang}, {Dai}, {Chen}, {Gao}, {Yang}, \& {Hao}}]{Zhou2024}
{Zhou}, L., {Wang}, T., {Daddi}, E., {et~al.} 2024, \aap, 684, A196

\end{thebibliography}

\clearpage

\begin{appendix}
\onecolumn

\clearpage
\section{Morphology}

\begin{figure}[!htbp]
    \centering
    \includegraphics[width=0.8\textwidth]{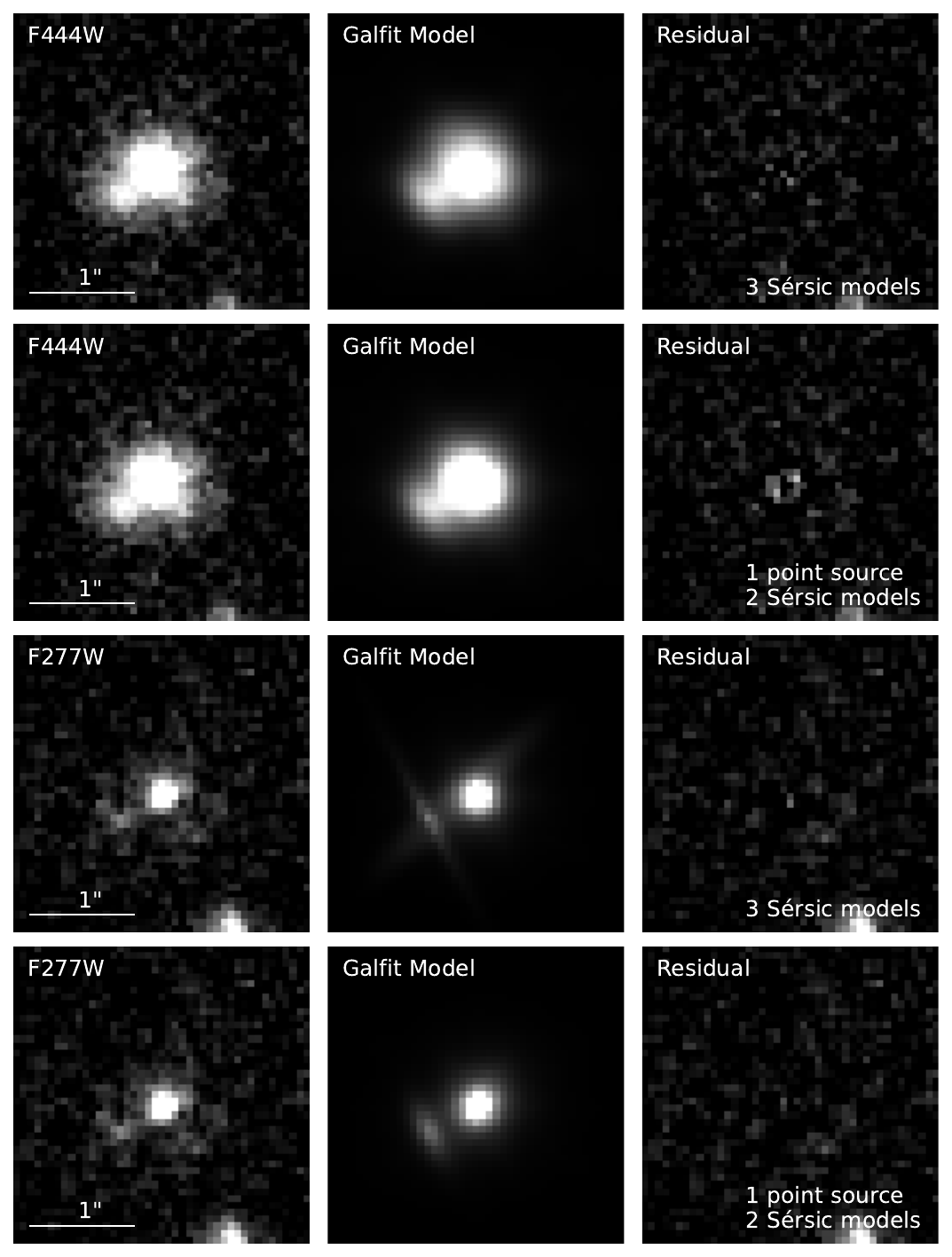}
    \caption{Morphological fit of XS55 in JWST/F444W (top and middle row) and JWST/F277W (bottom row). The different components of each fit are stated in the right column of each row.}
    \label{fig:morph}
\end{figure}

\twocolumn
\begin{table}[!htbp]
    \centering
    \renewcommand\arraystretch{1.5}
    \caption{Effective radii and Sérsic indices from Galfit}
    \begin{tabular}{c c}
       \multicolumn{2}{c}{\it F444W}\\
        \hline\hline
        \multicolumn{2}{c}{Compact component}\\
        $R_{\rm e,F444W}$ [kpc] & $0.40\pm0.05$ \\
        $n$ & $<1.2$\\
        \hline
        \multicolumn{2}{c}{Diffuse disk component}\\
        $R_{\rm e,F444W}$ [kpc] & $1.72\pm0.13$ \\
        $n$& $0.37\pm0.08$\\
        \hline
        \multicolumn{2}{c}{Companion}\\
        $R_{\rm e,F444W}$ [kpc] & $1.06\pm0.13$ \\
        $n$ & $0.61\pm0.33$\\
        \hline
        \multicolumn{2}{c}{F277W}\\
        \hline\hline
        \multicolumn{2}{c}{Compact component}\\
        $R_{\rm e,F277W}$ [kpc] & $0.58\pm0.07$ \\
        $n$ & $2.6\pm0.8$\\
        \hline
        \multicolumn{2}{c}{Diffuse disk component}\\
        \multicolumn{2}{c}{Component not detected}\\
        \hline
        \multicolumn{2}{c}{Companion}\\
        $R_{\rm e,F277W}$ [kpc] & $1.7\pm0.5$ \\
        $n$ & $2.0\pm1.4$\\
        \hline\hline \\
    \end{tabular}
    \label{tab:sizes}
\end{table}




\clearpage
\onecolumn
\section{FIR fitting}
\begin{figure*}[!htbp]
    \centering
    \includegraphics[width=0.49\textwidth]{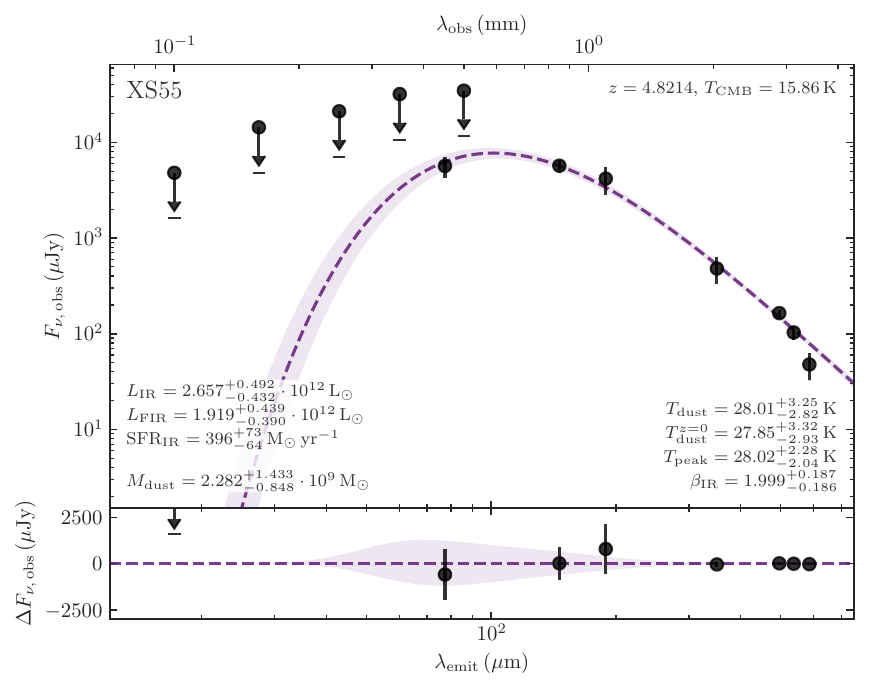}
    \includegraphics[width=0.49\textwidth]{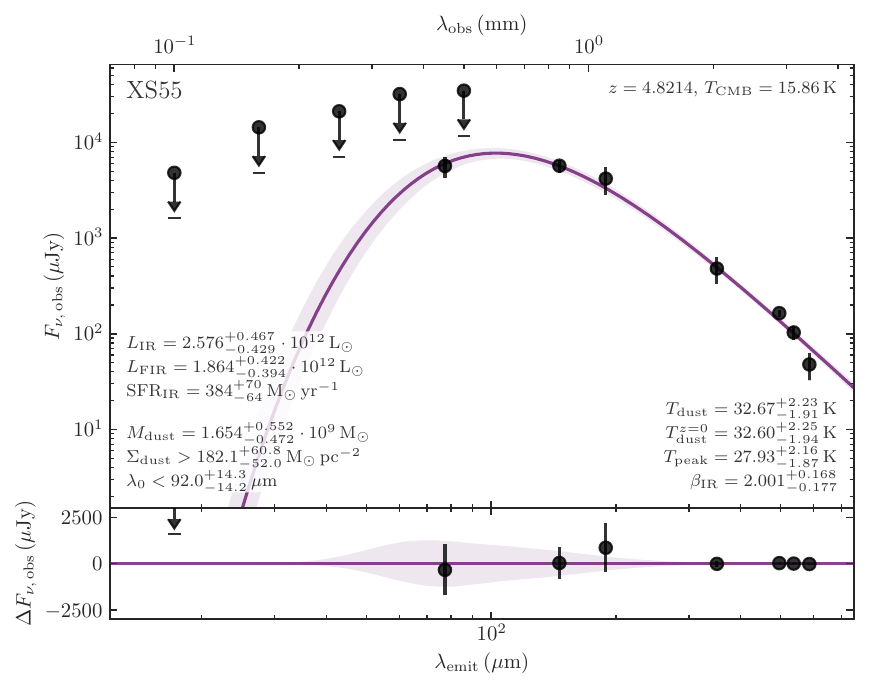}
    \caption{Mercurius FIR SEDs of XS55. {\bf Left:} FIR-SED assuming optically thin dust. {\bf Right:} FIR-SED assuming self-consistent optically thick dust, assuming an upper limit on the emission area of $9.08\,{\rm kpc^2}$ (half light radius of diffuse disk-like component in JWST/F444W). }
    \label{fig:FIR-SEDs}
\end{figure*}

\clearpage


\end{appendix}
\end{document}